\documentclass[manuscript]{aastex631}

\usepackage{esint}
\usepackage{appendix}
\usepackage{multirow}
\usepackage{makecell}
\usepackage{threeparttable}
\usepackage{longtable}
\usepackage[figuresright]{rotating}
\usepackage{footnote}
\usepackage{diagbox}
\usepackage{chngpage}
\graphicspath{{./}{figures/}}

\begin{document}

\title{Evolution of the Radial Size and Expansion of Coronal Mass Ejections Investigated by Combining Remote and In-Situ Observations}

\correspondingauthor{Bin Zhuang, No\'{e} Lugaz}
\email{bin.zhuang@unh.edu,noe.lugaz@unh.edu}

\author[0000-0002-5996-0693]{Bin Zhuang}
\affiliation{Institute for the Study of Earth, Oceans, and Space, University of New Hampshire, Durham, NH, USA}

\author[0000-0002-1890-6156]{No\'{e} Lugaz}
\affiliation{Institute for the Study of Earth, Oceans, and Space, University of New Hampshire, Durham, NH, USA}

\author[0000-0002-0973-2027]{Nada Al-Haddad}
\affiliation{Institute for the Study of Earth, Oceans, and Space, University of New Hampshire, Durham, NH, USA}

\author[0000-0002-9276-9487]{R\'eka~M. Winslow}
\affiliation{Institute for the Study of Earth, Oceans, and Space, University of New Hampshire, Durham, NH, USA}

\author[0000-0002-5681-0526]{Camilla Scolini}
\affiliation{Institute for the Study of Earth, Oceans, and Space, University of New Hampshire, Durham, NH, USA}

\author[0000-0001-8780-0673]{Charles J. Farrugia}
\affiliation{Institute for the Study of Earth, Oceans, and Space, University of New Hampshire, Durham, NH, USA}

\author[0000-0003-3752-5700]{Antoinette B. Galvin}
\affiliation{Institute for the Study of Earth, Oceans, and Space, University of New Hampshire, Durham, NH, USA}

\begin{abstract} A fundamental property of coronal mass ejections (CMEs) is their radial expansion, which determines the increase in the CME radial size and the decrease in the CME magnetic field strength as the CME propagates. CME radial expansion can be investigated either by using remote observations or by in-situ measurements based on multiple spacecraft in radial conjunction. However, there have been only few case studies combining both remote and in-situ observations. It is therefore unknown if the radial expansion estimated remotely in the corona is consistent with that estimated locally in the heliosphere. To address this question, we first select 22 CME events between the years 2010 and 2013, which were well observed by coronagraphs and by two or three spacecraft in radial conjunction. We use the graduated cylindrical shell model to estimate the radial size, radial expansion speed, and a measure of the dimensionless expansion parameter of CMEs in the corona. The same parameters and two additional measures of the radial-size increase and magnetic-field-strength decrease with heliocentric distance of CMEs based on in-situ measurements are also calculated. For most of the events, the CME radial size estimated by remote observations is inconsistent with the in-situ estimates. We further statistically analyze the correlations of these expansion parameters estimated using remote and in-situ observations, and discuss the potential reasons for the inconsistencies and their implications for the CME space weather forecasting.
\end{abstract}


\section{Introduction} \label{sec:intro}
Coronal mass ejections (CMEs) are large-scale solar eruptions that expel huge clouds of magnetized plasma and magnetic flux from the corona into interplanetary space. Radial expansion is one of the fundamental properties of CMEs, and always leads to an increase in the CME radial size and a decrease in the CME internal magnetic field strength with distance, which further affects the CME space weather effects \citep[e.g.,][]{gopal2014,gopal2015,lugaz2017b}. Studies on the radial expansion help to further understand the CME eruption process associated with magnetic reconnection \citep{zhuang2022}, the formation of the CME-driven shock \citep{patsourakos2010,lugaz2017b}, and the CME flux balance and erosion process \citep[e.g.,][]{dasso2006,ruffenach2012,lavraud2014} during the CME propagation. Over the past few decades, the CME radial expansion has been widely studied using (1) in-situ measurements which provide a time series of CME parameters (transformed into an one-dimensional (1-D) cut along the CME path) at a single point \citep[e.g.,][]{burlaga1982,farrugia1993,bothmer1998,liu2005,leitner2007,gulisano2010,winslow2016,good2019,vrsnak2019,lugaz2020,davies2020} and (2) remote observations which can be used to investigate the change of the CME morphology over time \citep[e.g.,][]{savani2009,lugaz2012,lugaz2020b,nieves2012,balmaceda2020,cremades2020}.

As for the in-situ measurements at a single spacecraft, one of the clearest signatures of the CME radial expansion is the decreasing profile of the CME bulk speed with time. This is used to estimate the expansion speed as half the front-to-back speed difference, which ranges from 10 to 250~km\,s$^{-1}$ \citep{burlaga1982,farrugia1993} with a typical value at 1~au of about 50~km\,s$^{-1}$. Such single-spacecraft measurements provide the CME local expansion properties. In order to study the global evolution of the CME radial expansion, statistics based on CMEs observed at different heliocentric distances have been used. Note that this does not require observing the same CME at multiple spacecraft. For example, combining Helios, Voyager and Pioneer~10, \citet{bothmer1998} found that between 0.3 to 4.2~au, the CME radial size follows, on average, a power law increase with heliocentric distance ($r_H$), $\propto r_H^{0.78}$. Similar statistical approaches were further made based on the data from Helios, Ulysses, Wind, and the Advanced Composition Explorer (ACE) \citep[e.g.,][]{liu2005,leitner2007,gulisano2010}. In general, those studies show that the radial size increases as $r_H^{0.6}$ to $r_H^{1.14}$ and the magnetic field strength decreases as $r_H^{-1.4}$ to $r_H^{-1.9}$ when considering only the measurements within 1~au or over a larger distance range of $\sim$0.3--6~au. Taking advantage of some planetary missions, e.g., using the data from the MErcury Surface, Space ENvironment, GEochemistry, and Ranging (MESSENGER) spacecraft, \citet{winslow2015} cataloged the CME events observed at MESSENGER and statistically found that from 0.3 to 1~au, the CME magnetic field strength decreases following $r_H^{-1.95}$. With the measurements from Juno, \citet{davies2021} revealed that the decrease in the CME magnetic field strength behaves differently inside and outside 1~au, and that the decrease slows down when the CME propagates to distances $>$1~au.

Multiple spacecraft in radial conjunction observing the same CMEs are helpful for our further understanding of the CME radial evolution. Since the beginning of solar cycle 24, the launch of the twin Solar Terrestrial Relations Observatory \citep[STEREO,][]{stereo} spacecraft, multiple planetary missions  (e.g., MESSENGER and Juno), and the recently launched Parker Solar Probe and Solar Orbiter, provide more opportunities for the same CMEs to be observed at different heliocentric distances with small longitudinal separations \citep[e.g.,][]{nieves2012,winslow2016,winslow2021a,wang2018,good2019,vrsnak2019,lugaz2020,salman2020,davies2020,davies2021,mostl2021}. For example, using a mapping technique to investigate the magnetic field time series of the same CME observed at different spacecraft, \citet{good2019} analyzed the underlying similarity of the magnetic structure of 18 CMEs in the inner heliosphere. They found that this similarity holds during the CME propagation. Based on the magnetic field evolution of 11 CME events in radial conjunction, \citet{vrsnak2019} found that the CME radial expansion is much stronger when CMEs are close to the Sun than at distances $>0.5$~au and deviates significantly from an isotropic self-similar expansion. Combining more samples reported in \citet{salman2020}, \citet{lugaz2020} found that the CME global expansion in the heliosphere estimated by the decrease in the CME magnetic field strength with distance is inconsistent with the local expansion property measured by the decrease in the CME bulk speed near 1~au. They also illustrated that different mechanisms might be responsible for the CME radial expansion in the innermost heliosphere and around 1~au.

Using remote observations, the CME expansion can be studied by the change of the CME morphology from the low corona by the Extreme-Ultra-Violet telescopes, to the middle and high corona by the white-light coronagraphs, and to interplanetary space by the Heliospheric Imagers (HIs) on board STEREO \citep[e.g.,][]{rouillard2009,savani2009,savani2010,nieves2012,veronig2018,balmaceda2020,cremades2020}. In the low corona, the CME lateral expansion is found to be faster than the radial one \citep{patsourakos2010,veronig2018}. Based on 475 CMEs observed in the STEREO coronagraphs, \citet{balmaceda2020} found that the average CME expansion speed is comparable to the average propagation speed at the CME center in the middle and high corona. Extending to STEREO~HIs with a wider field-of-view (FOV), the CME expansion can be tracked farther in interplanetary space. For example, \citet{savani2009} continuously tracked a CME in HI-1 images and found that its radial size increases obeying a power law with heliocentric distance ($\propto r_H^{0.6}$) between 0.1 and 0.4~au. We note that this exponential index lies at the bottom of the statistical range described above, which indicates a slower radial expansion for this event. \citet{nieves2012} further combined remote and radially aligned in-situ observations covering a distance of 1~au to investigate the expansion evolution both in radial and lateral directions of one CME case. They found that the CME morphology at 1~au reconstructed based on the in-situ measurements is consistent with the reconstruction based on remote observations. Case studies on the CME radial expansion properties combining both remote and in-situ observations were also done by, e.g., \citet{nieves2013} and \citet{lugaz2020b}.

Measurements with multiple spacecraft in radial conjunction were not a frequent occurrence until solar cycle 24, e.g., there were about 50 events between the years 2011 and 2014 reported by \citet{salman2020}. It is even rarer for the same event to be continuously tracked using remote observations (especially by STEREO~HIs within a wider FOV) and multiple in-situ instruments in radial conjunction. We note again the investigation of one particularly well-observed CME event, which was associated with a consistent radial size as estimated remotely and in situ \citep{nieves2012}. However, there has been no statistical investigation of the CME radial expansion combining both remote and radially aligned in-situ observations, and it is still unknown whether all CMEs have a consistent radial expansion from the corona to interplanetary space as found for one event in \citet{nieves2012}. It has been proposed that, in the innermost heliosphere, CMEs expand due to their high internal magnetic pressure; while reaching 1~au the expansion is due to the decrease in the solar-wind total pressure with distance \citep[e.g.,][]{demoulin2009,lugaz2020}. These different mechanisms driving the CME radial expansion at different distances may lead to the inconsistency of the radial expansion estimated remotely in the corona (or close to the Sun) and in situ in interplanetary space, e.g. near 1~au. In this paper, we aim to statistically study the CME radial expansion and provide general properties by combining remote and radially aligned in-situ observations. We note that, for remote observations, only coronagraphs are used instead of taking into account the STEREO/HIs images, because (1) only few events in our sample can be well observed by STEREO/HIs, and (2) the CME exact boundaries (e.g., leading edge and trailing edge) in the HIs FOV are always difficult to identify.

The rest of the paper is organized as follows. We describe the data, event selection, calculation of the expansion parameters, and analysis methods in Section~\ref{sec: data}. In Section~\ref{sec: comp1}, we present the observations and detailed analyses for a CME, which erupted on 2013 July 9, and, then, show the comparison of the radial sizes obtained using the remote and in-situ observations for the total 22 CME events. Section~\ref{sec: comp2} shows the statistics of the comparisons of other expansion parameters. Discussion and conclusions are given in Sections~\ref{sec: dis} and \ref{sec: con}.

\section{Data and Methods} \label{sec: data}
\subsection{CME Event Selection} \label{sec: database}
We start from the list of 47 CME events observed in radial conjunction between the years 2011 and 2015 as listed in \citet{salman2020} and \citet{lugaz2020}. These include the events measured by MESSENGER \citep{messenger} orbiting between $\sim$0.31 and $\sim$0.47~au, Venus Express \citep[VEX;][]{vex} at $\sim$0.72--0.73~au, the twin STEREO spacecraft, Wind, and ACE \citep{ace} near 1~au. Note that only magnetic field measurements are used for the planetary missions MESSENGER and VEX \citep{salman2020,lugaz2020}, while both magnetic field and solar wind plasma measurements are available for the spacecraft near 1~au. Events in this database are measured by two spacecraft with a longitudinal separation of less than 35$^\circ$. More details can be found in \citet{salman2020} and \citet{lugaz2020}. 

In this paper, we additionally require that: (1) the CME can be clearly observed in coronagraphs and the CME shape can be well captured by the graduated cylindrical shell model in order to  perform the related CME 3-D reconstruction (Section~\ref{sec: quanti2}), (2) the CME propagation direction is close to the ecliptic plane, and (3) there is no CME-CME interaction. The consideration of the second criterion is to minimize the influence of the CME deflection in latitude \citep[e.g.,][]{shen2011,kay2015,mostl2015}. The fact that CME-CME interaction can result in changes in the CME kinematics and internal magnetic properties \citep[e.g.,][]{lugaz2012,lugaz2017a} leads to the incorporation of the third criterion. 19 events from the original database meet these three additional criteria. Furthermore, we also include three more events that occurred during MESSENGER's cruise phase with the CME eruption time on 2010 June 16 \citep[also see][]{nieves2012}, 2010 November 3, and 2010 December 12. The heliocentric distances of MESSENGER during these times were 0.56~au, 0.47~au, and 0.36~au, respectively.  Overall, there are 14 events having a MESSENGER-1~au conjunction, and 9 events having a VEX-1~au conjunction. The CME that erupted on 2011 November 3 was the only event with a MESSENGER-VEX-1~au conjunction.

The in-situ boundaries of the CMEs are obtained from \citet{winslow2015} for MESSENGER, from \citet{good2016} for VEX, from \citet{richardson2010} and \citet{chi2016} for ACE and Wind, and from \citet{jian2018} for STEREO, with two exceptions. For the CMEs measured at MESSENGER and VEX, the boundaries were checked visually to ensure that the corresponding magnetic field profiles at these two spacecraft are consistent with those measured near 1~au: (1) the front boundary at MESSENGER for the 2013-July-9 CME was revised to 04:05 UT on July 11 \citep[also see][]{lugaz2020b}, and (2) the rear boundary at VEX for the 2011-March-16 CME was revised to 22:00 UT on March 19. Table \ref{cme_table} lists the information about the selected 22 CMEs, including the expansion parameters described in the next two subsections.

\subsection{Methods for In-situ Measurements}\label{sec: quanti1}
To study the expansion properties of CMEs based on in-situ measurements, we calculate the CME radial size ($S_r$), radial expansion speed ($V_{\rm{exp}}$), center speed ($V_{\rm center}$), a dimensionless expansion parameter ($\zeta$), and two parameters indicating the decrease in the magnetic field strength ($\alpha_B$) and the increase in the radial size ($\alpha_r$) with heliocentric distance ($r_H$). In the following part, the term ``CME'' for the in-situ measurements refers to the magnetic ejecta counterpart.
		
We begin with two calculations of $V_{\rm{exp}}$ as done in \citet{salman2020} and \citet{lugaz2020}. The first calculation takes the entire CME period into consideration, and thus $V_{\rm{exp}}$ is derived as:
\begin{equation}\label{eq1}
        V_{\rm{exp}}=(V_{\rm{front}}-V_{\rm{back}})/2,
\end{equation}
where $V_{\rm{front}}$ and $V_{\rm{back}}$ are the bulk speeds estimated at the CME front and rear boundaries, respectively. The second calculation only considers a limited period where a nearly linear decrease in the speed profile inside the CME is present. Within this period, a slope of $\Delta V / \Delta t$ is obtained. This slope is applied to the entire CME period to predict the CME bulk speed variation, and the corresponding expansion speed is then derived using Equation \ref{eq1}, following \citet{gulisano2010}. Examples of the second calculation can be found in Figure~\ref{fig: 2013july}(a), Figure~14 in \citet{salman2020}, and Figure~2 in \citet{lugaz2020}. The expansion speeds obtained based on the first and second methods are hereafter denoted as $V_{\rm{exp-meas}}$ and $V_{\rm{exp-fit}}$.

We calculate the CME radial size with and without expansion corrected, and the latter calculation provides an upper limit of the size value. The radial size without expansion corrected is calculated as follows:
\begin{equation}\label{eq2}
	S_r=V_{\rm{impact}} \times \Delta t,
\end{equation}		
where $V_{\rm{impact}}$ is the CME impact speed (or, its initial estimate) at a spacecraft, and $\Delta t$ is the in-situ CME duration. The radial size with radial expansion speed corrected is estimated as:
\begin{equation}\label{eq3}
	S_r'=(V_{\rm{impact}}-V_{\rm{exp}}) \times \Delta t.
\end{equation}	
Near 1~au, we use (a) $V_{\rm{exp-fit}}$ which is more appropriate to some cases with a disturbed speed profile, and (b) the maximum CME speed as $V_{\rm{impact}}$ \citep{salman2020}. For most of the CMEs studied here, the speed maximum is located near the front boundary. 

The estimation of $V_{\rm{impact}}$ and $V_{\rm{exp}}$ at MESSENGER and VEX is indirect due to the lack of plasma measurements. Following \citet{salman2020}, $V_{\rm{impact}}$ is calculated by a three-step measurement process based on the drag-based model \citep[DBM,][]{vrsnak2013} that considers the acceleration or deceleration of CMEs due to the CME-solar wind interaction in interplanetary space. This process first adjusts the drag coefficient in the DBM to make the prediction match (1) the CME arrival time near 1~au, (2) $V_{\rm{impact}}$ estimated near 1~au, and (3) the CME arrival time at MESSENGER or VEX. $V_{\rm{impact}}$ at MESSENGER or VEX is then derived by averaging the three predicted impact speeds and the corresponding error is the standard deviation. In DBM, the inner boundary is set as 20~$R_\odot$, and the CME initial speed is the quadratically-fitted CME speed at 20~$R_\odot$ obtained from the Coordinated Data Analysis Workshop's (CDAW) CME catalog \citep{yashiro2004}. The calculation of $V_{\rm{exp}}$ assumes two conditions: (a) the same expansion speed as that estimated near 1~au, and (b) a linearly decreasing one with distance from 20~$R_\odot$ to $\sim$1~au ($V_{\rm{exp}}$ at 20~$R_\odot$ is introduced in Section~\ref{sec: quanti2}). The uncertainty in the radial size is obtained by combining the $V_{\rm{impact}}$ error estimates as well as the different estimates for $V_{\rm{exp}}$.

The dimensionless parameter $\zeta$ provides a local measure of the CME radial expansion and is used to deal with the problem that larger and faster CMEs tend to have larger expansion speeds \citep{demoulin2009,gulisano2010,lugaz2020}. $\zeta$ is defined as: 
\begin{equation}\label{eq4}
	\zeta = \frac{r_H}{V_{\rm center} ^2} \frac{\Delta V}{\Delta t} \sim \frac{r_H}{S_r}\frac{2V_{\rm{exp}}}{V_{\rm center}},
\end{equation}
where $V_{\rm center}$ is the average speed within the CME period. The use of $V_{\rm{exp-meas}}$ and $V_{\rm{exp-fit}}$ lead to $\zeta_{\rm{meas}}$ and $\zeta_{\rm{fit}}$, respectively. The two calculations of $\zeta$ are equivalent only for the cases where the speed can be linearly fitted over the entire CME period. More details are given in \citet{lugaz2020}.

Parameters $\alpha_B$ and $\alpha_r$ are defined as:
\begin{equation}\label{eq5}
	\alpha_B=\frac{\log (B_2/B_1)}{\log (r_{H2}/r_{H1})}, \ {\rm{and}} \ \alpha_r=\frac{\log (S_{r2}/S_{r1})}{\log (r_{H2}/r_{H1})},
\end{equation}
where $B$ is the CME magnetic field strength. Subscripts 1 and 2 refer to the first (inner, i.e., MESSENGER or VEX) and second (outer, i.e., STEREO-A, STEREO-B, or the spacecraft at L1) spacecraft at different heliocentric distances. Both $\alpha_{Bavg}$ and $\alpha_{Bmax}$ based on the average ($B_{avg}$) and maximum ($B_{max}$) magnetic field strengths inside CMEs are calculated \citep[see][]{lugaz2020}. In general, $\alpha_r$ and $\alpha_B$ represent a global expansion behavior of CMEs \citep{dumbovic2018,salman2020,lugaz2020,scolini2021}.
 
\subsection{Methods for Remote Observations}\label{sec: quanti2}
We combine remote observations from multiple viewpoints of the Large Angle and Spectrometric Coronagraph on board the SOlar and Heliospheric Observatory \citep[SOHO/LASCO;][]{lasco} and the coronagraphs COR1 and COR2 \citep{secchi} on board STEREO-A and STEREO-B. To further eliminate the projection effects and obtain more precise CME expansion parameters, we use the graduated cylindrical shell (GCS) model \citep{thernisien2006,thernisien2009} that assumes a flux-rope shape and a self-similar expansion to reconstruct the CME morphology in 3-D space. There are six free parameters used in the GCS model, which are the height of the CME leading edge ($h$), latitude and longitude of the CME propagation direction ($\theta$ and $\phi$), tilt angle ($\gamma$), aspect ratio ($\kappa$), and the CME angular half width ($\alpha$). When reconstructing the CME at different time steps, only $h$ is adjusted and the remaining five parameters are fixed.

Since there is no direct magnetic field measurement for CMEs in the corona, useful parameters revealing the CME expansion properties based on the GCS model are the radial size $S_r$, radial expansion speed $V_{\rm{exp}}$, and the dimensionless parameter $\zeta$. Here we describe the calculation processes in detail. We calculate the parameters along the CME propagation direction first: (1) the radius of the cross-section of the flux rope is $R=\frac{\kappa h}{1+\kappa}$, and thus $S_r=2R$, indicative of the maximum radial size; (2) $V_{\rm{exp}}$ is derived by the linear fit of $R$ versus time; (3) the CME center speed $V_{\rm center}$ is derived by the linear fit of ($h-R$) versus time; (4) $\zeta_{\rm{GCS}} \sim \frac{r_H}{S_r}\frac{2V_{\rm{exp}}}{V_{\rm center}}=\frac{h}{2R}\frac{2V_{\rm{exp}}}{V_{\rm center}}=1+\kappa$. These derived parameters are a function of $\kappa$, and a fixed $\kappa$ requires that the CME expands self-similarly as it propagates. We do not use a time-varying $\kappa$ in the GCS model, as it may add additional uncertainties due to the fact that identifying the CME trailing edge can be difficult in remote observations (discussed in Section~\ref{sec: dis}). As such, we keep the self-similar expansion assumption and $\kappa$ constant.

In Sections~\ref{sec: 20130709} and \ref{sec: cmp_radial_size}, the CME propagation direction in latitude is found to be close to the ecliptic plane for most of the CME events. Therefore, the calculations of $V_{\rm{exp}}$ and $\zeta_{\rm{GCS}}$ along the propagation direction are used to represent the CME expansion properties in the corona for distances $\le 20$~$R_\odot$. Besides,  $\alpha_r$ is not derived for the GCS model, because it equals unity theoretically: $\alpha_r=\frac{\log (S_{r2}/S_{r1})}{\log (r_{H2}/r_{H1})}=\frac{\log (\frac{\kappa h_2}{1+\kappa}/\frac{\kappa h_1}{1+\kappa})}{\log (h_2/h_1)}$=1. We estimate $S_r$ in the ecliptic plane along the middle longitude between the two spacecraft. To further study the evolution of the CME radial size in interplanetary space based on the GCS model, we linearly fit $S_r$ versus $r_H$ in the logarithmic scale of the CME leading edge in the ecliptic plane and along the Sun-spacecraft line, and then extrapolate $S_r$ outward using the fitted results. Further details are given in Section~\ref{sec: 20130709}.

The uncertainties in the parameters of the GCS model are discussed here. As shown in \citet{thernisien2009}, the average uncertainties are $0.48$~$R_\odot$ for $h$, $4.3^\circ$ for $\phi$, $0.9^\circ$ for $\theta$, $22^\circ$ for $\gamma$, $0.07$ for $\kappa$, and $13^\circ$ for $\alpha$, respectively. To simplify the calculations of the radial expansion parameters, we only consider the effects of the height ($\Delta h$) and aspect ratio ($\Delta \kappa$). This consideration is reasonable because (1) the uncertainties of some of the remaining four parameters, namely the propagation direction, are relatively small, and (2) their effects on the CME radial size can be even minimal when the CME propagates roughly in the ecliptic plane and close to the Sun-spacecraft line which is true for most of the selected CMEs. The uncertainty in $S_r$ equals $2\times \Delta R=2\times \sqrt{ (\frac{\kappa \Delta h}{1+\kappa})^2 + (\frac{h\Delta \kappa}{(1+\kappa)^2})^2 }$, and the uncertainty in $\zeta$ is 0.07 since $\zeta=1+\kappa$. The uncertainty in $V_{\rm exp}$ refers to the 1-$\sigma$ error in the linear fit, which ranges from around 10 to 80~km~s$^{-1}$ for $V_{\rm exp}$ in the range of around 45 to 355~km~s$^{-1}$ for our events.

In order to compare the radial size estimated using remote and in-situ observations, we adopt Equation \ref{eq6} to calculate the corresponding difference: 
\begin{equation}\label{eq6}
	{\rm{dif}}=\frac{S_{{r-\rm{insitu}}}-S_{{r-\rm{GCS}}}}{S_{{r-\rm{GCS}}}}\times 100\%,
\end{equation}
where $S_{r-\rm{insitu}}$ is the radial size estimated in situ, and $S_{r-\rm{GCS}}$ is the extrapolated GCS result at the distance of an in-situ spacecraft. The error in the size difference is estimated by Equation~\ref{eq7}:
\begin{equation}\label{eq7}
	{\delta \rm{dif}}=\sqrt{(\frac{\delta S_{r-\rm{insitu}}}{S_{r-\rm{GCS}}})^2 + (\frac{S_{r-\rm{insitu}} \delta S_{r-\rm{GCS}}}{S_{r-\rm{GCS}}^2})^2} \times 100\%,
\end{equation}
where $\delta S_{r-\rm{insitu}}$ and $\delta S_{r-\rm{GCS}}$ are the uncertainties in the radial size estimates. When the difference by using the average of the in-situ sizes estimated with and without correcting for the radial expansion speed is within $\pm 20 \%$, we consider that the two estimates of the radial size are consistent. We note that the value of $20\%$ is set empirically and was purposefully chosen to be relatively strict by considering the uncertainties in the radial size estimates (see Table~\ref{tab: radial_size_tb}), and a slight modification of this value does not significantly affect the group categorization of the radial size consistency (also see Figure~\ref{fig: radial_dif}).

\subsection{Statistical Methods}\label{sec: sta}
To estimate the correlation between two sample populations, we use (a) the linear Pearson correlation that assesses a linear relationship between two variables and (b) the Spearman's rank correlation assessing monotonic (whether linear or not) relationships. The associated correlation coefficients are denoted $\rho_P$ and $\rho_S$ hereafter. Before calculating the corresponding coefficient of a pair of variables, we use the Shapiro-Wilk test to check whether the individual variable follows a normal distribution (95\% confidence level). If the normality is confirmed, both correlation methods are used; otherwise, only the Spearman's rank correlation is estimated. We then calculate the uncertainties in correlation coefficients as follows: (1) using the bootstrapping resampling method with replacement to select 22 (the total event number) sample pairs (the same pair could be repetitively selected) and calculating the related correlation coefficient, (2) iterating the first step for a large number of times (5000 in this paper) and obtaining a coefficient group including 5000 data points, and (3) using the 1-$\sigma$ error with a 68\% confidence level as the uncertainty in the coefficient value. We note that the small sample size and large uncertainties in the measurements limit the use of the 2-$\sigma$ error (the corresponding error values are comparable to or larger than the coefficient values for most pairs of parameters), and thus we focus on the 1-$\sigma$ error in this paper and it is not possible to obtain results at the 95\% (2-$\sigma$) confidence level. Furthermore, we use the nonparametric Wilcoxon test and calculate its effect size to check if the radial size differences at different distances have substantially different averages.

\section{Comparison of CME Radial Size} \label{sec: comp1}
In this section, we first show the observations and detailed analyses of one CME event that started on 2013 July 9. We then present the comparison of the radial size estimated using remote and in-situ observations for the 22 events.

\subsection{2013-July-9 CME Event}\label{sec: 20130709}
The CME erupted on 2013 July 9 and was subsequently measured in situ by MESSENGER and Wind. At this time, the longitudinal separation between the two spacecraft was $\sim$3$^\circ$, and MESSENGER was at the heliocentric distance of $\sim$0.45~au. This event was also studied by \citet{lugaz2020b} who focused on the long duration of the CME and its sheath region. They raised a question about the origin of the long-duration ejecta, as the CME is expanding relatively slowly at 1~au, typically between MESSENGER and 1~au and is already wide at MESSENGER. Here, we can partially address this question by investigating the radial expansion of this CME in the corona.

\begin{figure}[!hbt]
    \centering
    \includegraphics[width=\textwidth]{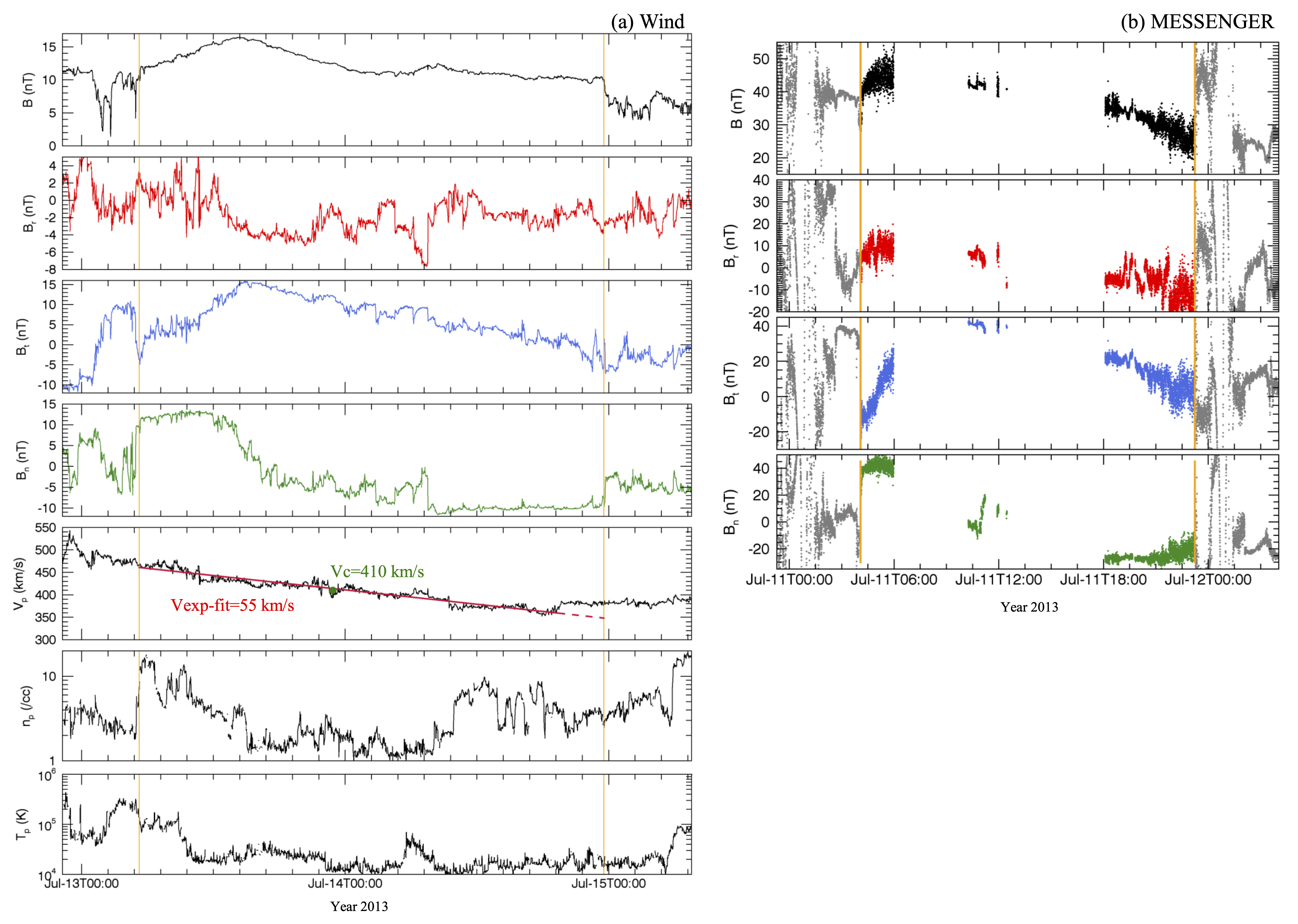}
    \caption{\small (a) Wind measurements showing, from top to bottom, the magnetic field strength and components in RTN coordinates, the solar wind proton bulk speed ($V_p$), number density ($n_p$), and temperature ($T_p$). The red line indicates the linearly fitted results for estimating the CME expansion speed, and the CME center speed is marked by the green dot. (b) MESSENGER measurements of the magnetic field strength and components in RTN coordinates. The vertical orange lines in the two panels indicate the in-situ magnetic ejecta boundaries. In panel (b), the magnetic field measurements outside the CME region are plotted in gray, and the time periods when MESSENGER was inside Mercury's magnetosphere are removed.}
    \label{fig: 2013july}
\end{figure}
Figure~\ref{fig: 2013july} shows the in-situ measurements at Wind (a) and MESSENGER (b). Panel~(a) from top to bottom shows the Wind measurements of the magnetic field strength (black), components in the spacecraft-centered Radial Tangential Normal (RTN) coordinates, solar wind proton bulk speed ($V_p$), number density ($n_p$), and temperature ($T_p$). The CME boundaries are marked by the vertical orange lines. This event shows typical in-situ characteristics of CMEs, namely: (1) an increase in the magnetic field strength and a decrease in the proton temperature indicative of a magnetically-dominated structure, (2) a smooth rotation of the magnetic field component, and (3) a linear decrease in proton bulk speed indicative of a CME radial expansion \citep[e.g.,][]{burlaga1982,richardson2010}. The red line in the velocity profile shows the linear fit to the data for the expansion speed estimation using the second method described in Section \ref{sec: quanti1}. The solid line corresponds to the region with a linearly decreasing trend, and the dashed part is the extension of the linear fit to the remaining region. The CME center speed is indicated by the green dot in the figure. 

Panel~(b) from top to bottom shows the MESSENGER observations of the magnetic field strength and components in RTN coordinates, in which the CME boundaries are marked by the orange lines. Inside the CME region, the time periods when MESSENGER was inside Mercury's magnetosphere have been removed. The time evolution of the magnetic field profile at MESSENGER is quite consistent with that at Wind. The duration of the CME is found to be around 20 hours at MESSENGER, and around 42 hours at 1~au. Furthermore, based on the Wind plasma measurements, $V_{\rm{exp-meas}}$ is 42~km~s$^{-1}$, $V_{\rm{exp-fit}}$ is 55~km~s$^{-1}$, $V_{\rm center}$ is 410~km~s$^{-1}$, $\zeta_{\rm{meas}}$ is 0.49, and $\zeta_{\rm{fit}}$ is 0.64, indicating that the radial expansion estimated in situ is smaller than average for CMEs at 1~au \citep{lugaz2020b}.

\begin{figure}[!hbt]
	\centering
	\includegraphics[width=0.9\textwidth]{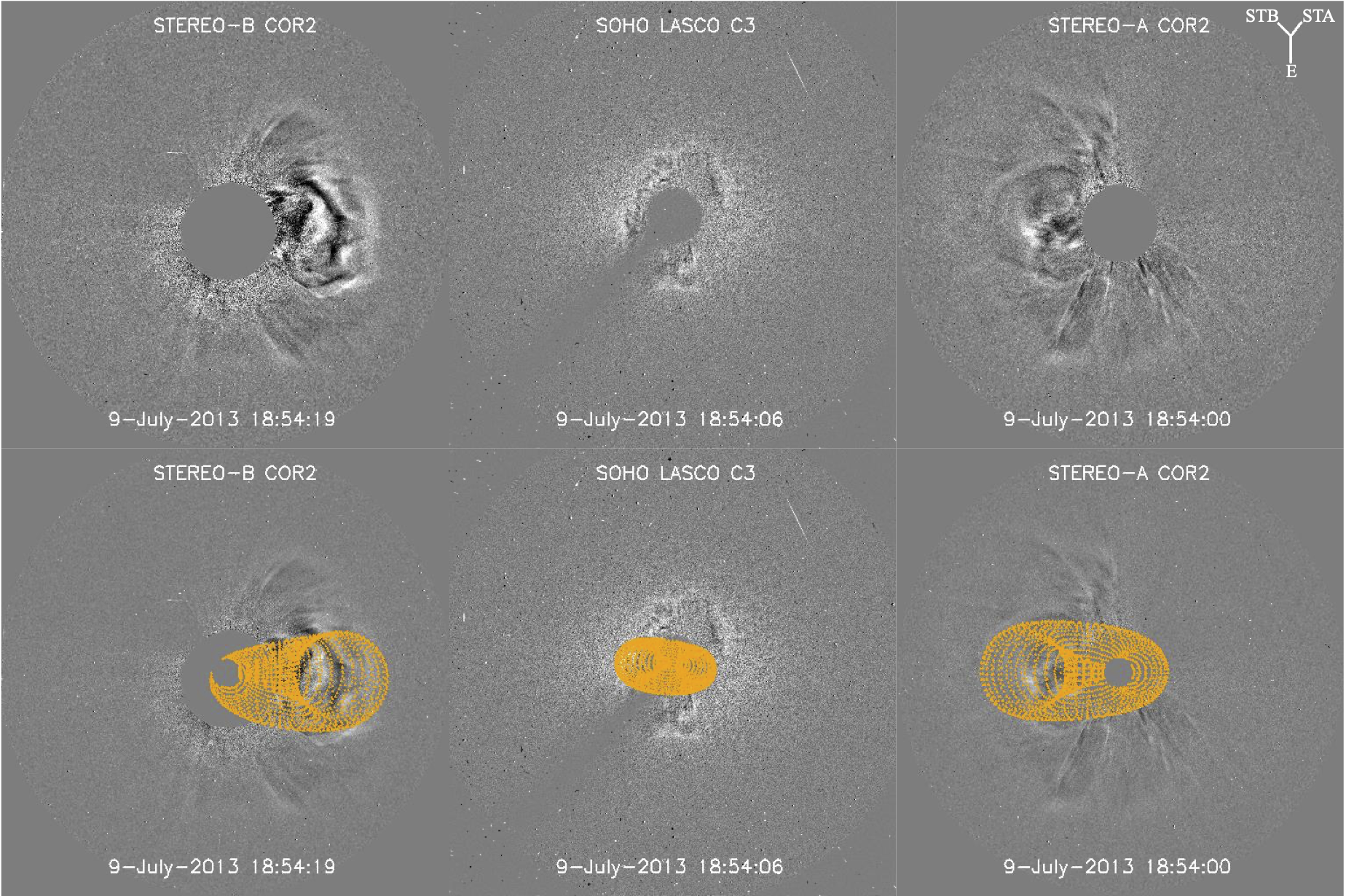}
	\caption{\small Reconstruction of the CME on 2013 July 9 using the GCS model. Top: running-difference images of the CME observed nearly simultaneously at 18:54 UT on 2013 July 9 by STEREO-B/COR2 (left), SOHO/LASCO/C3 (middle), and STEREO-A/COR2 (right). The insert in the top right panel shows the locations of STEREO-B, Earth, and STEREO-A relative to the Sun. Bottom: reconstructed flux rope structure (orange) overlapped.}
	\label{fig: gcs}
\end{figure}

We then investigate the CME radial expansion in the corona using the GCS model. Figure \ref{fig: gcs} shows the reconstruction of the CME on 2013 July 9. The upper panels are the running-difference images of the CME observed nearly simultaneously in STEREO-B/COR2 (left), SOHO/LASCO/C3 (middle), and STEREO-A/COR2 (right). The insert in the top right panel shows the locations of STEREO-B, Earth, and STEREO-A relative to the Sun.
The lower panels have the reconstructed flux rope structure (orange) overlapped. Based on the GCS model, the CME propagates along N02E16, with the morphology parameters of $\kappa \sim 0.36$ and $\alpha=45^\circ$, and $V_{\rm center}$ and $V_{\rm{exp}}$ of 346 and 127~km~s$^{-1}$, respectively. $\zeta_{\rm{GCS}}$ is thus calculated as 1.36.

\begin{figure}[!hbt]
    \centering
    \includegraphics[width=\textwidth]{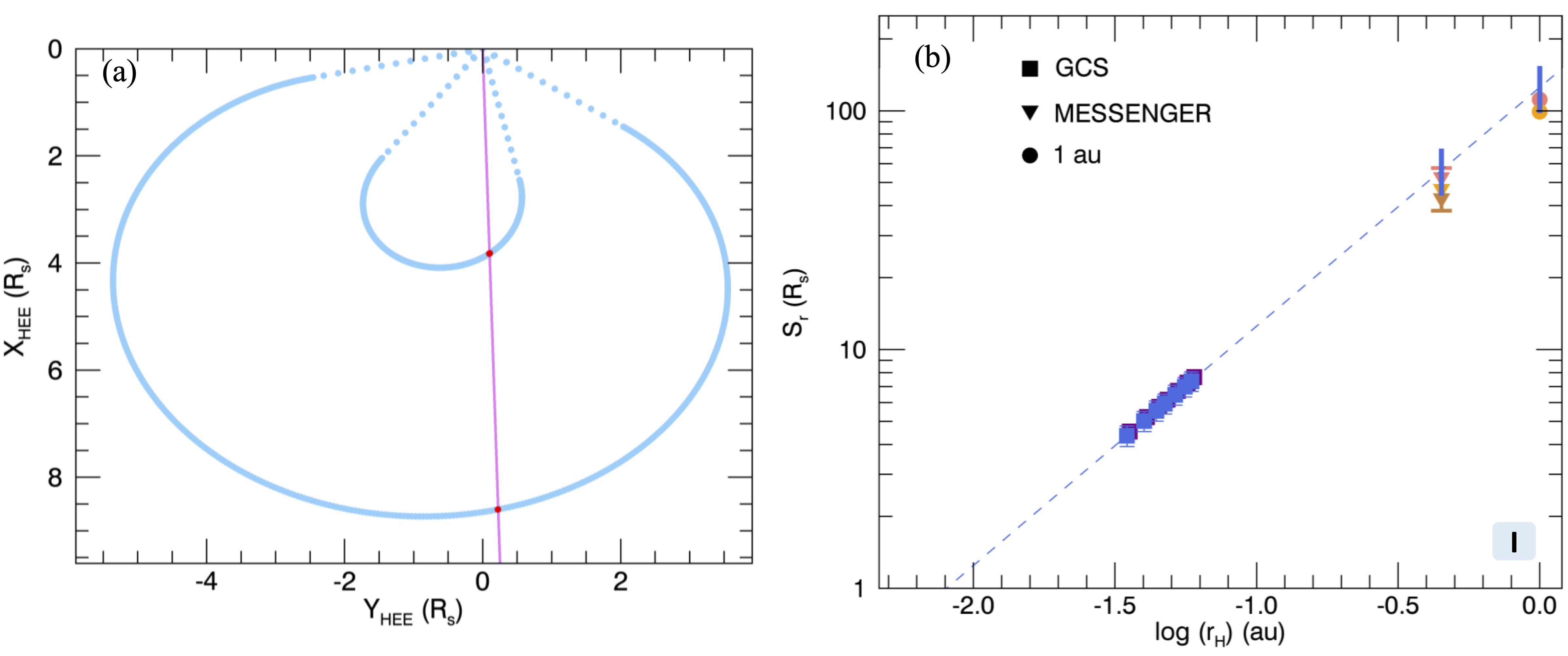}
    \caption{\small (a) CME cross-section in the ecliptic plane at 16:39~UT on 2013 July 9 based on the GCS model. The magenta line indicates the middle longitude between Wind and MESSENGER. (b) Variation of the CME radial size along with heliocentric distance estimated using the GCS model (square) with error bars overlaid, and using the in-situ measurements at MESSENGER (downward triangle) and Wind (circle). The uppermost and lowermost endcaps of the error bar at MESSENGER correspond to the two extreme uncertainties by incorporating the $V_{\rm impact}$ error. The dashed blue line indicates the linear fit to the data points of $\log S_r$ versus $\log r_H$, and the $\pm 20\%$ uncertainties of the GCS radial size at MESSENGER's distance and 1~au are shown by two vertical bars. The consistency category is listed by a roman numeral in the bottom right corner. See more information in the text.}
\label{fig: gcs_ecliptic_insitu}
\end{figure}

In order to ensure a consistent comparison with the in-situ measurements at the spacecraft roughly orbiting in the ecliptic plane, we obtain the GCS flux-rope shape and parameters intersected in the same plane as shown in Figure~\ref{fig: gcs_ecliptic_insitu}(a). The magenta line in this panel indicates the middle longitude between Wind and MESSENGER, and we measure the radial size of the CME (bounded by the two red dots) along this line.
Figure~\ref{fig: gcs_ecliptic_insitu}(b) shows the evolution of the CME radial size along with heliocentric distance estimated using the GCS model (square symbol) and in-situ measurements (triangles and circles). The blue square shows the CME radial size in the ecliptic plane, and the distance corresponds to the leading edge of the intersected cross-section along the purple line as shown in Figure~\ref{fig: gcs_ecliptic_insitu}(a). Error bars of the CME radial size estimation in the corona are plotted, and the size uncertainties of $\pm 20\%$ in the GCS model are also indicated by two vertical blue bars at MESSENGER's distance and 1~au. 

We use a linear fit applied to the data points of $\log S_r$ versus $\log r_H$ (dashed blue line) to estimate the radial size of the CME in interplanetary space extrapolated from the GCS model. The purple square shows the radial size along its propagation direction in 3-D space. For this event, the blue and purple data points almost overlap. At MESSENGER, the top triangle shows the estimated CME radial size without the radial expansion considered, and the middle and bottom ones show the radial sizes with the expansion speed corrected based on the first and second estimation methods as described in Section \ref{sec: quanti1}. The uppermost and lowermost error-bar endcaps correspond to the two extreme uncertainties in the radial size estimation after incorporating the error of $V_{\rm impact}$ (the same as those in Figures~\ref{fig: radial_exp_p1} to \ref{fig: radial_exp_p3}). At Wind, the top and bottom circles correspond to the sizes without and with the radial expansion considered, respectively. The CME radial size estimated from the in-situ measurements is found to be smaller than that from remote observations estimated using the GCS model, by $-7\%$ to $-24\%$ at MESSENGER without and with correcting for the radial expansion, and by $-12\%$ to $-21\%$ at 1~au without and with considering the expansion, respectively. Using the consistency threshold of $\pm 20\%$ (Section~\ref{sec: quanti2}), the radial size estimated in situ is in agreement with that estimated remotely when considering the averages of the radial size differences. 

We come back to the long duration of this CME measured in situ. The expansion speed in the corona obtained from the GCS model is not large, compared to other events with magnitudes of $\sim$200--300~km\,s$^{-1}$ as shown in Figure \ref{fig: cmp_glo_ins}(b). However, the large fitted $\kappa$ in this event leads to a large CME radial size itself and a higher $\zeta_{\rm{GCS}}$. These two factors may play a role in the long duration of the CME during its propagation in the heliosphere, while the role of $\zeta_{\rm{GCS}}$ in the CME radial expansion is discussed in Section \ref{sec: comp2}. 

\subsection{Radial Size Evolution from the Corona to $\sim$1~au: Statistical Results}\label{sec: cmp_radial_size}
\begin{figure}[!hbt]
    \centering
    \includegraphics[width=0.92\textwidth]{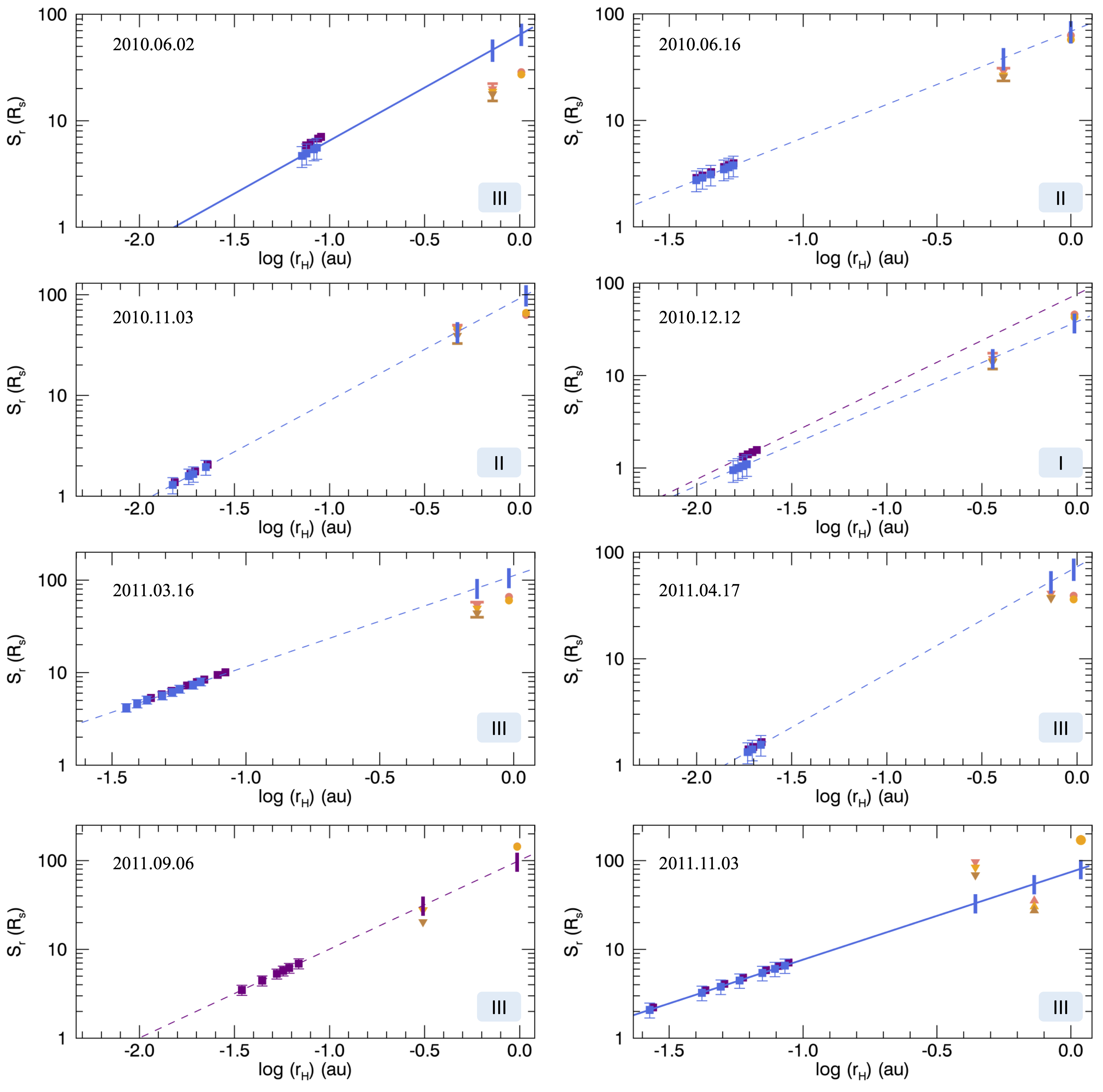}
    \caption{\small Evolution of the CME radial size estimated using remote and in-situ observations. The squares show the GCS model results (purple: along the propagation direction in 3-D space; blue: in the ecliptic plane and along the Sun-spacecraft line) with error bars overlaid, the triangles show the in-situ estimations at MESSENGER or VEX, and the circles show the estimations at STEREO or Wind/ACE. The in-situ estimations are the results with and without the CME expansion speed considered in different colors. The dashed line indicates the linear fit to the squares, and the vertical bars at the distances of the in-situ spacecraft indicate the $\pm 20\%$ differences of the radial size from the GCS model. The two endcaps for some events at MESSENGER or VEX indicate the uppermost and lowermost uncertainties by incorporating the $V_{\rm impact}$ errors. The consistency category of each event is listed in the bottom right corner with a roman numeral. See more details in Section \ref{sec: 20130709}.}
    \label{fig: radial_exp_p1}
\end{figure}

\begin{figure}[!hbt]
	\centering
	\includegraphics[width=0.92\textwidth]{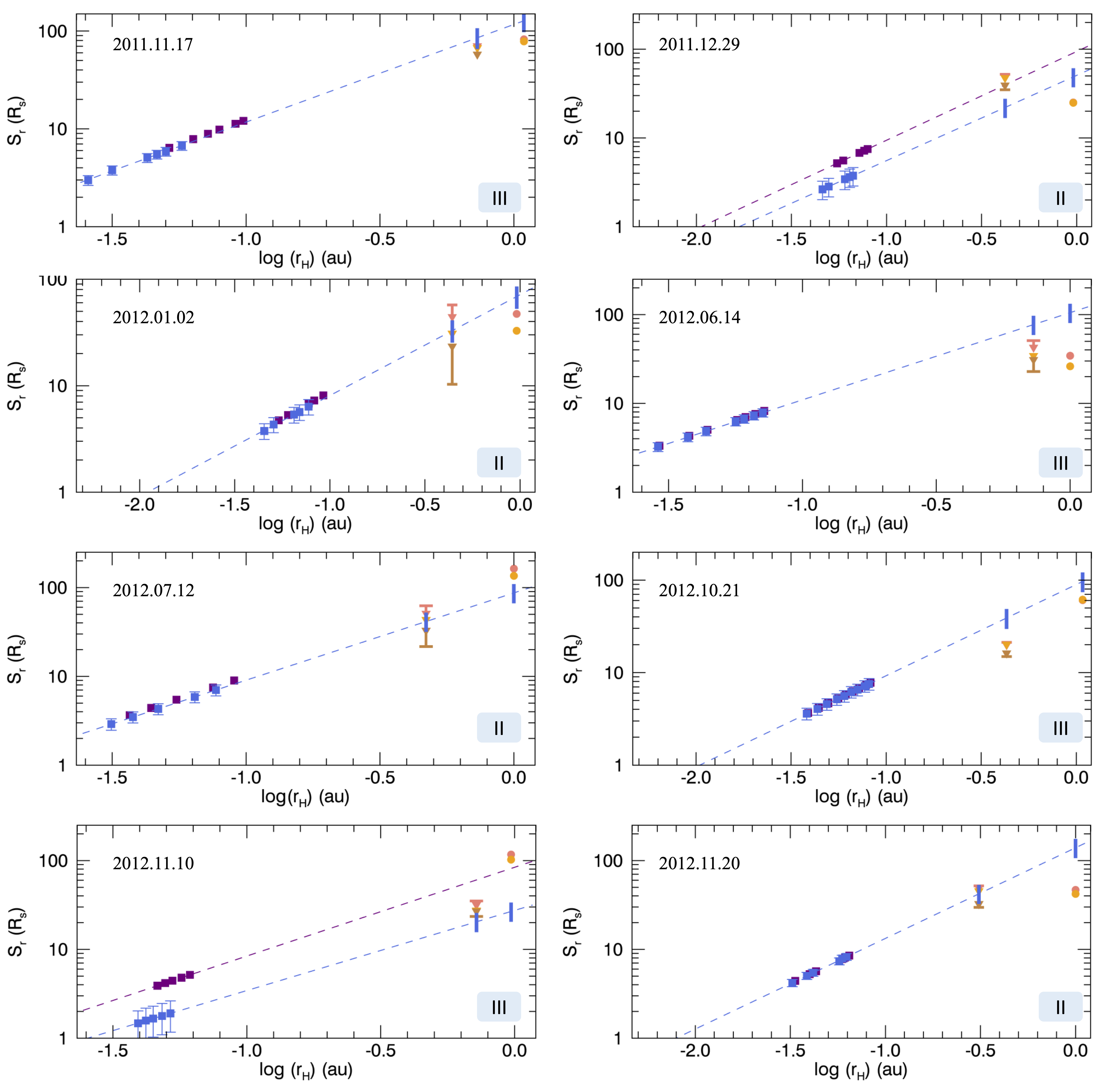}
	\caption{\small Same as Figure \ref{fig: radial_exp_p1}.}
	\label{fig: radial_exp_p2}
\end{figure}

\begin{figure}[!hbt]
	\centering
	\includegraphics[width=0.92\textwidth]{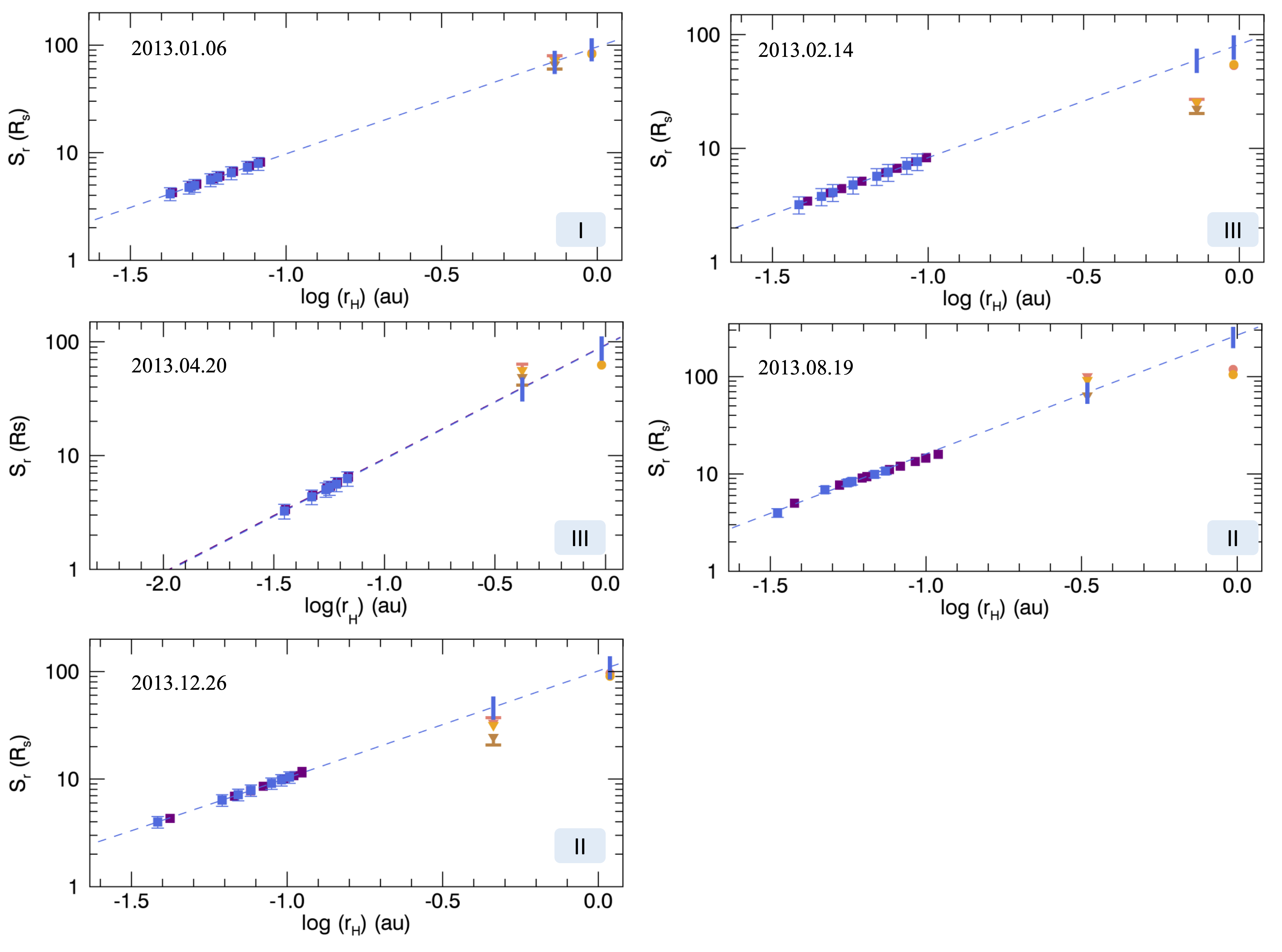}
	\caption{\small Same as Figure \ref{fig: radial_exp_p1}.}
	\label{fig: radial_exp_p3}
\end{figure}

Figures~\ref{fig: radial_exp_p1} to \ref{fig: radial_exp_p3} show the evolution of the CME radial size estimated using the GCS model in remote observations and in-situ measurements at different heliocentric distances for the remaining 21 CME events (chronologically sequenced). In these figures, the blue data points are close to the purple data points for most events due to the fact that the CMEs propagate close to the ecliptic plane. There is thus no linear fit to the purple squares for those events. If the blue and purple data points are substantially different, we also linearly fit the purple ones. We note that there are no blue squares for the event on 2011 September 6 because of the relatively higher fitted latitude of the CME propagation direction in the GCS model. Besides, for some events there are no endcaps of the errors of the size estimates at MESSENGER or VEX by considering the $V_{\rm impact}$ error because the drag coefficients in the three-step process are constant and thus the standard deviation of $V_{\rm impact}$ is zero. The CME event started on 2011 November 3 in radial conjunction with MESSENGER-VEX-1~au was previously studied by \citet{good2015,good2018} and \citet{salman2020}. In Figure~\ref{fig: radial_exp_p2}, the extrapolation of the radial size along the CME propagation direction (dashed purple line) is consistent with the in-situ measurements at MESSENGER for the event on 2011 December 29 and at STEREO-A for the event on 2012 November 10. It may be associated with the fact that the CME still experiences a deflection in latitude.

We turn our attention to the consistency of the CME radial size estimates between remote and in-situ observations for the remaining 21 events, with the associated radial sizes, size differences, and their uncertainties estimated remotely and in situ as listed in Table~\ref{tab: radial_size_tb}. For the 2011-December-29 CME at MESSENGER and the 2012-November-10 CME at STEREO-A, we use the GCS extrapolation along the propagation direction (purple square), for the other events we use the blue data points. We note that the consistency of the radial size estimation is based on an average of the in-situ radial sizes calculated with and without expansion speed corrected. Therefore, the condition of the GCS $\pm20\%$ error bars overlapping only one of the in-situ estimates (e.g., the one without the expansion speed corrected) does not mean a true consistency. We categorize the events into three groups. Group I includes events for which the evolution of the radial size from the GCS model matches the radial sizes measured at both in-situ spacecraft. There are only three such CME events (2010 December 12, 2013 January 6, and 2013 July 9). Group II includes the CMEs for which the evolution of the CME radial size from the GCS model matches that obtained at one in-situ spacecraft only. Eight events (2010 June 16, 2010 November 3, 2011 December 29, 2012 January 2, 2012 July 12, 2012 November 20, 2013 August 19, and 2013 December 26) are in this group. Group III includes the events for which the radial sizes are in disagreement between the GCS model and in-situ estimates, and is composed of 11 events (2010 June 2, 2011 March 16, 2011 April 17, 2011 September 6, 2011 November 3, 2011 November 17, 2012 June 14, 2012 October 21, 2012 November 10, 2013 February 14, and 2013 April 20). The group categorization for each event is shown in Figures \ref{fig: gcs_ecliptic_insitu} to \ref{fig: radial_exp_p3} and listed in Table \ref{tab: consistency}. For group II, the radial size derived from the remote observations are consistent with that from the in-situ measurements at MESSENGER for six out of eight events and near 1~au for the remaining two events. For the event on 2010 June 16, the differences are $\sim$$-26\%$ to $-30\%$ at MESSENGER and $\sim$$-13\%$ to $-20\%$ at 1~au without and with expansion corrected, which may still indicate a quite good agreement between the remote and in-situ estimations, as discussed by \citet{nieves2012}.

\begin{figure}[!hbt]
	\centering
	\includegraphics[width=0.9\textwidth]{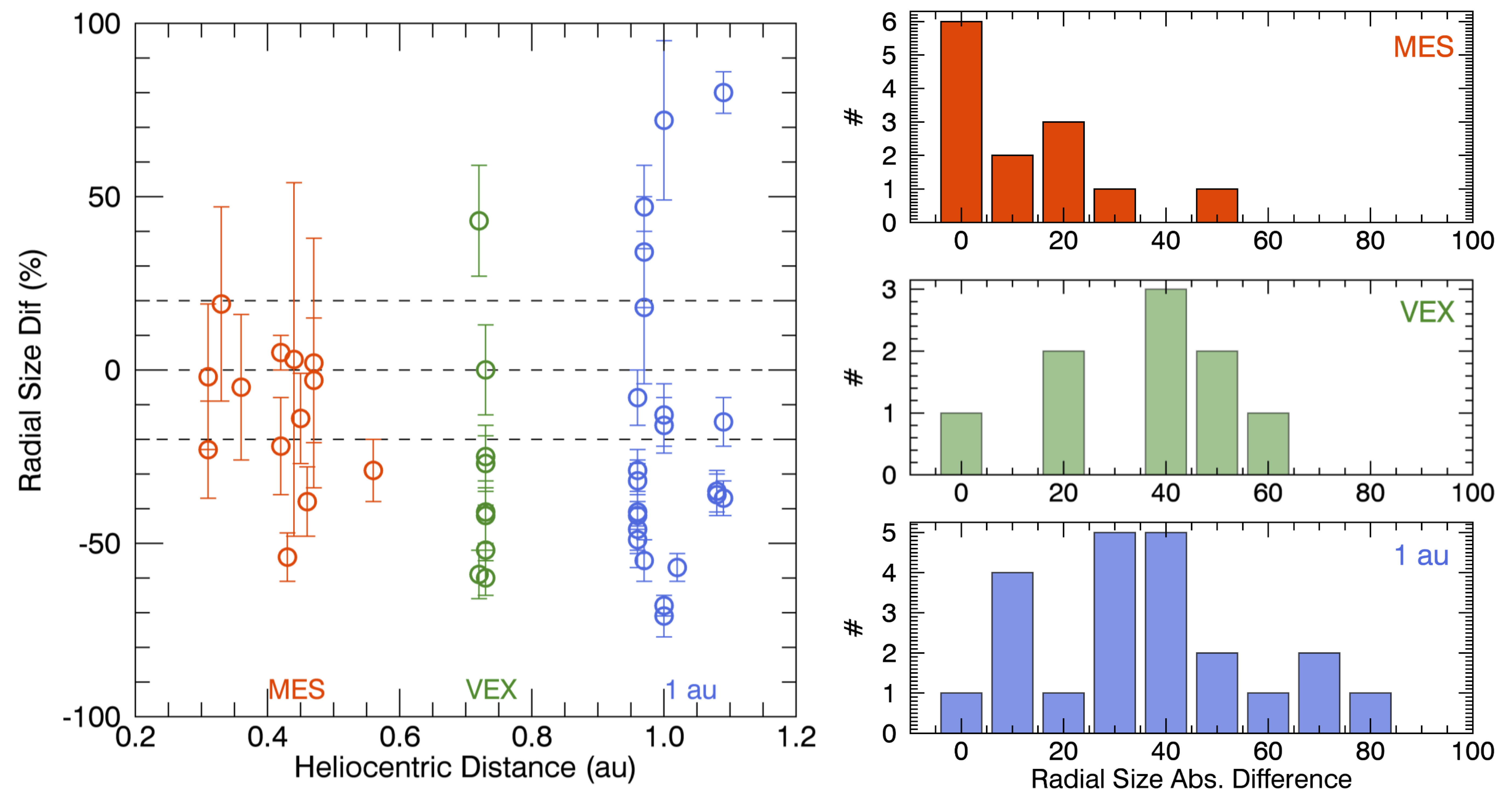}
	\caption{\small Left panel: difference of the radial size estimated by Equation \ref{eq6} as a function of heliocentric distance and the associated uncertainty estimated by Equation~\ref{eq7}. Red, green, and blue refer to MESSENGER, VEX, and spacecraft near 1~au, respectively. An outlier of 150$\%$ for MESSENGER at 0.44~au for the 2011-November-3 event is not shown. The two horizontal dashed lines indicate the $\pm 20\%$ differences. Right panels: histograms of the absolute value of the radial size difference estimated at MESSENGER, VEX, and the spacecraft near 1~au.}
	\label{fig: radial_dif}
\end{figure}
The left panel of Figure \ref{fig: radial_dif} shows the difference of the radial size estimated using Equation~\ref{eq6} and the uncertainty in the size difference estimated by Equation~\ref{eq7} at different distances for all the events, in which an outlier of 150$\%$ at MESSENGER (with large uncertainty) for the 2011-November-3 event is not considered and plotted. It is found that, for all spacecraft of MESSENGER (red), VEX (green), and spacecraft near 1~au (blue), the radial size estimated from remote observations using the GCS model, in general, overestimates the radial size based on in-situ measurements, as is clear from the fact that there are more data points distributed below zero. This is further confirmed when considering the averages of the size difference: $-12\% \ \pm 20\%$ at MESSENGER, $-29\% \ \pm 33\%$ at VEX, and $-18\% \ (\pm 43\%)$ at the spacecraft near 1~au. However, the average of the absolute size difference at MESSENGER ($16\% \ \pm 16\%$) is found to be lower than those at VEX ($39\% \ \pm 19\%$) and at spacecraft near 1~au ($41\% \ \pm 20\%$). The nonparametric Wilcoxon test further confirms that the average of the absolute radial size difference estimated at MESSENGER is significantly lower than those estimated at VEX and the spacecraft near 1~au according to the corresponding probability values being lower than the significance level of 0.05 (0.02 for MESSENGER versus VEX, and 0.001 for MESSENGER versus the spacecraft near 1~au) with effect sizes greater than 0.5 (1.1 for MESSENGER versus VEX and 1.4 for MESSENGER versus the spacecraft near 1~au). The three right panels of Figure~\ref{fig: radial_dif} show the histograms of the absolute value of the radial size difference estimated at MESSENGER, VEX, and the spacecraft near 1~au, which further indicate that the absolute size difference estimated at MESSENGER is smaller than those estimated at VEX and the spacecraft near 1~au. In addition, out of the 14 CME events which have measurements at MESSENGER, eight events have consistent radial sizes between the remote (GCS) and in-situ (at MESSENGER) estimates. The results above indicate that CMEs may have a consistent radial expansion from the corona to the innermost heliosphere (e.g., within heliocentric distances of $<$0.5~au). Table \ref{tab: size_dif} summarizes the averages of the (absolute) radial size differences associated with their 1-$\sigma$ uncertainties.

\section{Comparison of the CME Radial Expansion Parameters}\label{sec: comp2}
Table \ref{tab: cc} lists the correlation coefficients, $\rho_P$ and $\rho_S$ between the parameters estimated from remote observations using the GCS model and from the in-situ measurements. The 1-$\sigma$ uncertainties are estimated using the bootstrapping resampling method as described in Section~\ref{sec: sta}. The GCS parameters include $V_{\rm exp}$, $\zeta_{\rm GCS}$, $V_{\rm center}$, and $V_{\rm front}$, where $V_{\rm front}=V_{\rm center}+V_{\rm exp}$ is the speed of the CME front (or leading edge) along the propagation direction. The in-situ parameters include $V_{\rm{exp-fit}}$, $V_{\rm{exp-meas}}$, $V_{\rm center}$, $\zeta_{\rm{fit}}$, and $\zeta_{\rm{meas}}$ near 1~au, the radial size ($S_r$, the average of the values with and without the radial expansion speed corrected) at MESSENGER, VEX, and the spacecraft near 1~au, as well as the global expansion parameters of $\alpha_{Bmax}$, $\alpha_{Bavg}$, and $\alpha_{r}$. Based on the Shapiro-Wilk test, (a) all parameters estimated in the GCS model, (b) the radial sizes estimated at MESSENGER, VEX, and the spacecraft near 1~au, and (c) $V_{\rm{exp-meas}}$, $V_{\rm center}$, and $\zeta_{\rm{meas}}$ estimated using the in-situ measurements near 1~au follow a normal distribution. Thus, their $\rho_P$ values are also calculated. This paper focuses on the radial expansion properties, and thus the comparisons between $V_{\rm center}$ and $V_{\rm front}$ in the GCS model and those in-situ expansion parameters are not discussed in detail. We note that it is quite common for CMEs with high coronal speeds to have stronger expansions \citep{schwenn2005,gopalswamy2009,balmaceda2020} and to continue propagating fast to $\sim$1~au \citep[e.g.,][]{mostl2014}.

\subsection{$V_{\rm{exp}}$ and $\zeta$}\label{sec: cmp_para1}
\begin{figure}[!hbt]
    \centering
    \includegraphics[width=0.95\textwidth]{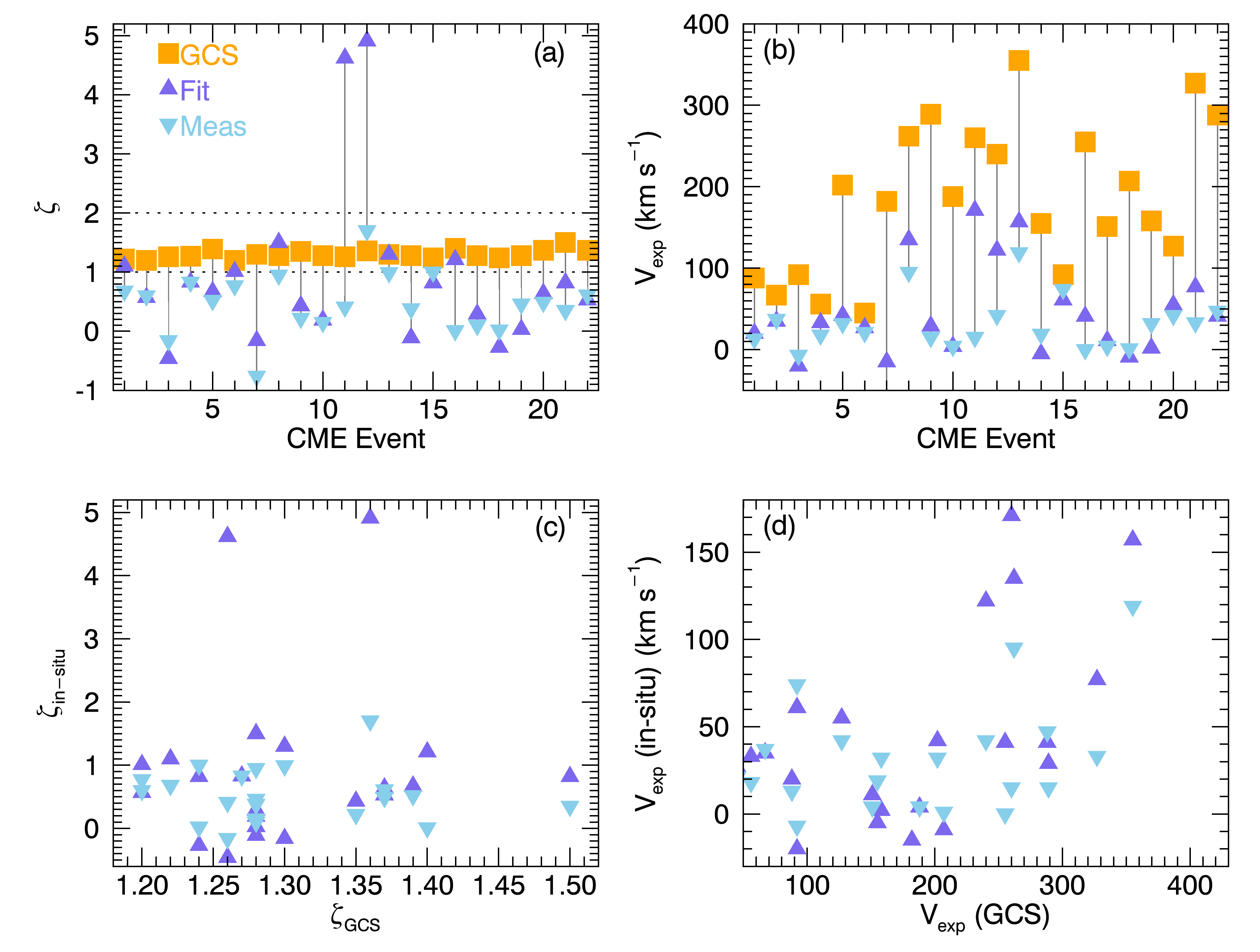}
    \caption{\small Comparison of the CME radial expansion parameters of $\zeta$ and $V_{\rm{exp}}$ estimated using the GCS model and in-situ measurements near 1~au. The orange square shows the results from the GCS model, and the purple upward and blue downward triangles indicate the in-situ parameters derived by the fitting and measuring as described in Section~\ref{sec: quanti1}, respectively.}
    \label{fig: cmp_glo_ins}
\end{figure}

Figure~\ref{fig: cmp_glo_ins} shows the comparisons of $V_{\rm{exp}}$ and $\zeta$ obtained from remote observations and in-situ measurements near 1~au. As described in Section \ref{sec: quanti2}, $\zeta_{\rm{GCS}}$ only depends on the fitted $\kappa$, which leads to $\zeta_{\rm{GCS}}$ ranging in $\sim$1.2--1.5 as shown by the orange square in panel~(a). In order to indicate an extreme condition of $\zeta_{\rm{GCS}}$ and compare that with the in-situ parameters, we manually set a low limit of $\zeta_{\rm{GCS}}$ to be 1 and an upper limit to be 2. The upward triangles indicate $\zeta_{\rm{fit}}$ and $V_{\rm{exp-fit}}$, while the downward ones refer to $\zeta_{\rm{meas}}$ and $V_{\rm{exp-meas}}$. Panel~(b) shows the comparison of the radial expansion speeds estimated remotely and in situ. We found that (1) both $\zeta_{\rm{fit}}$ and $\zeta_{\rm{meas}}$ are lower than $\zeta_{\rm{GCS}}$ for most of the events, and (2) $V_{\rm{exp}}$ estimated in situ is smaller than $V_{\rm{exp}}$ estimated by the GCS model, which is consistent with CMEs having a stronger expansion in the corona as compared to that near 1~au. 

Panels~(c) and (d) show the correlations between $\zeta$ and $V_{\rm{exp}}$ estimated remotely and in situ. We look at the expansion speed first. $V_{\rm{exp}}$ in the GCS model moderately correlates with $V_{\rm{exp-fit}}$ near 1~au with $\rho_S$ of $0.46 \pm 0.17$. $V_{\rm{exp-meas}}$ is not discussed due to the weak correlations ($\rho \lesssim 0.3 \ \pm0.2$). Comparing the values of $\zeta$ obtained using the GCS model and in-situ estimates, there is no correlation as the corresponding $\rho$ values are close to 0. We recall Equation~\ref{eq4} that $\zeta$ estimated by the in-situ measurements near 1~au depends on a combination of $S_r$, $V_{\rm{exp-fit}}$ or $V_{\rm{exp-meas}}$, and $V_{\rm center}$. Based on the moderate correlations between $V{\rm{_{exp-GCS}}}$-$V{\rm{_{exp-fit}}}$ and $V{\rm{_{exp}}}$-$V_{\rm center}$ ($\rho \sim 0.45 \ \pm 0.2$ as shown in Table~\ref{tab: cc}), the disagreement between $\zeta_{\rm GCS}$ and the in-situ $\zeta$ may be due to the inconsistencies of the radial size, as presented in Section~\ref{sec: cmp_radial_size}. In addition, we find that there exists a weak correlation between $\zeta_{\rm GCS}$ and $V_{\rm{exp-fit}}$ with $\rho_S$ to be 0.32 $\pm 0.17$.

\subsection{$V_{\rm{exp}}$ and $\zeta$ in the GCS Model versus $S_r$}
We further study the correlations between $V_{\rm{exp}}$ and $\zeta$ in the GCS model versus $S_r$ (averaging the values with and without expansion corrected) estimated at MESSENGER, VEX, and near 1~au. Based on Table~\ref{tab: cc}, there exist weak correlations between $V_{\rm{exp}}$ in the GCS model and $S_r$ estimated near 1~au and at MESSENGER. The corresponding $\rho_P$ and $\rho_S$ increase from $0.38 \pm 0.18$ and $0.35 \pm 0.20$ near 1~au (22 events) to $0.49 \pm 0.19$ and $0.46 \pm 0.24$ at MESSENGER (14 events). However, such a correlation is not clear at VEX because the uncertainties in $\rho_P$ and $\rho_S$ are comparable to or larger than the coefficient values. Higher $\rho_P$ and $\rho_S$ estimated at MESSENGER compared to those estimated at the spacecraft near 1~au are consistent with the result described in Section~\ref{sec: cmp_radial_size}: the absolute difference of the radial size at MESSENGER is smaller than that estimated near 1~au. It further indicates that the CME initial expansion in the corona plays a role in the evolution of the CME radial size (as also described in Introduction), which is dominant in the innermost heliosphere. 

Between $\zeta_{\rm{GCS}}$ and $S_r$ estimated near 1~au, a weak correlation exists with $\rho_S$ to be $0.36 \pm 0.19$ ($\rho_P$ is not used due to the relatively larger uncertainty). The correlation between $\zeta_{\rm{GCS}}$ and $S_r$ becomes stronger at MESSENGER and VEX. The corresponding $\rho_P$ and $\rho_S$ are $0.46 \pm 0.28$ and $0.38 \pm 0.23$ at MESSENGER and $0.52 \pm 0.22$ and $0.52 \pm 0.29$ at VEX, respectively. It indicates that $V_{\rm center}$ may also play a role in the CME radial expansion as the CME propagates in the heliosphere. In general, the CME radial size is roughly proportional to the radial expansion speed and inversely proportional to the propagation speed, while the later form determines the length of the time that a CME expands. This may lead to a quite good correlation between $\zeta_{\rm{GCS}}$ and $S_r$, as also shown in \citet{gulisano2010} where $S_r$ $\propto r^\zeta$. If a CME propagates to, e.g., $>$0.7~au, we may expect that its propagation speed reaches the solar wind speed due to the solar wind drag force \citep{vrsnak2013}, and thus the correlation between $\zeta_{\rm{GCS}}$ and $S_r$ may become weaker.

\begin{figure}[!hbt]
    \centering
    \includegraphics[width=0.95\textwidth]{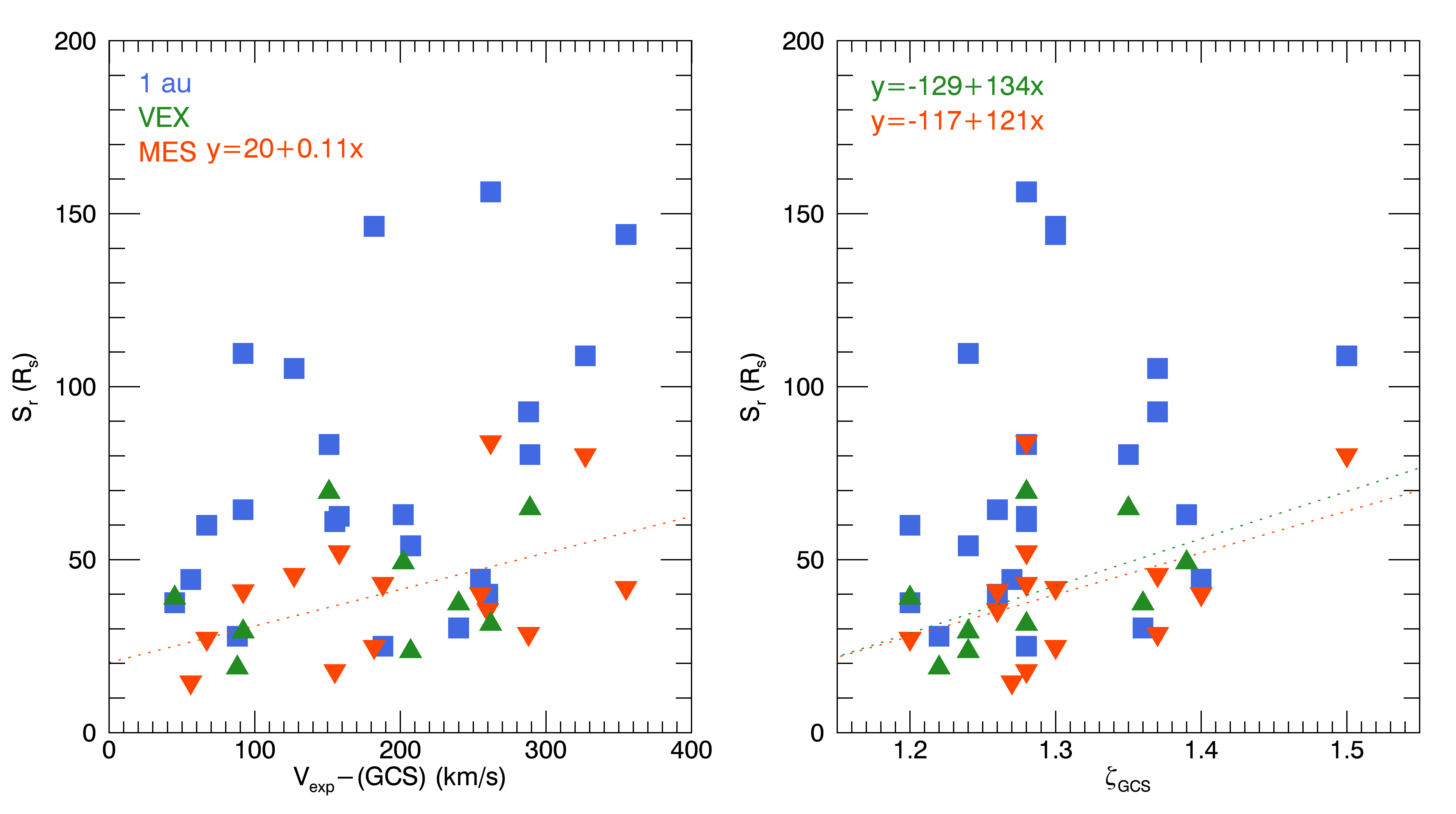}
    \caption{\small Correlations between $V_{\rm{exp}}$ from the GCS model and the in-situ $S_r$ (left panel) and between $\zeta_{\rm{GCS}}$ and $S_r$ (right panel). The data points at different spacecraft are plotted in different colors (MESSENGER: red; VEX: green; near 1-au spacecraft: blue). The dotted line indicates the linear fit to the data points only with their correlation coefficients $>0.4$.}
    \label{fig: gcs_insitu_size}
\end{figure}
Figure~\ref{fig: gcs_insitu_size} shows the correlations between $V_{\rm{exp}}$ in the GCS model and $S_r$ estimated in situ at different spacecraft in the left panel, and between $\zeta_{\rm{GCS}}$ and $S_r$ in the right panel. Data points at different spacecraft are plotted in different colors (MESSENGER: red; VEX: green; 1~au spacecraft: blue). The dotted line indicates the linear fit to the data points only with correlation coefficients (either $\rho_P$ or $\rho_S$) exceeding $0.4$. As for the $\zeta_{\rm{GCS}}$-$S_r$ relationship, the linear fit to the data points at VEX has a similar slope compared to that at MESSENGER (see the results in the figure). It may indicate that the dependence of the evolution of the CME radial size on $\zeta_{\rm{GCS}}$ is not significantly changed when the CME propagates from the Sun to interplanetary space at least within $\sim$0.7~au.

\subsection{$\alpha_B$ and $\alpha_r$}\label{sec: cmp_other_para}
\begin{figure}[!hbt]
    \centering
    \includegraphics[width=\textwidth]{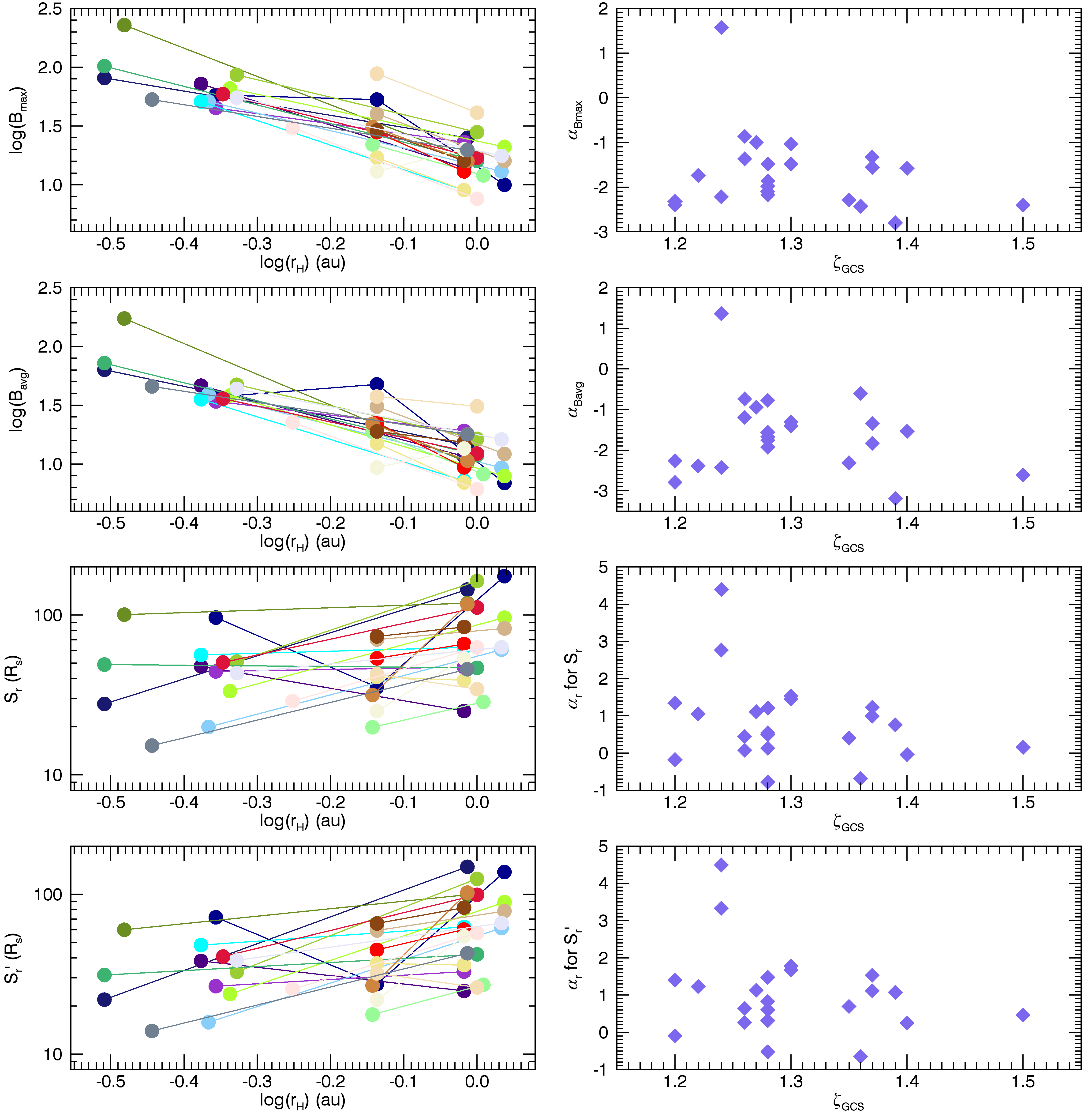}
    \caption{\small Comparisons of the CME radial expansion parameters. The left panels show the evolution of the CME magnetic field strength (maximum and average) and radial size (without, and with, the correction for the expansion speed) with heliocentric distance. The right panels show $\zeta_{\rm{GCS}}$ versus $\alpha_B$ and $\alpha_r$.}
    \label{fig: cmp_exp_para}
\end{figure}
Figure \ref{fig: cmp_exp_para} shows the comparisons of the parameters of $\alpha_B$ and $\alpha_r$ indicative of the CME global expansion in interplanetary space based on the radially aligned in-situ measurements versus $\zeta_{\rm{GCS}}$ indicative of the CME expansion in the corona. \citet{lugaz2020} found that the CME global radial expansion indicated by $\alpha_B$ and $\alpha_r$ is not consistent with the local expansion ($\zeta$) estimated in situ at the spacecraft near 1~au. The left panels of Figure~\ref{fig: cmp_exp_para} show the estimations of the global expansion parameters, i.e., $\alpha_{Bmax}$, $\alpha_{Bavg}$, and $\alpha_r$ as introduced in Section \ref{sec: quanti1}. Similar figures for the magnetic field information can be found in some past studies \citep[e.g.,][]{salman2020,lugaz2020,scolini2021}. In this figure, $S_r$ and $S'_r$ correspond to the radial size estimates without, and with, the correction for the radial expansion speed, respectively. At MESSENGER and VEX, we use the greater $V_{\rm{exp}}$ estimated by the second method as described in Section~\ref{sec: quanti1}. 

For the selected 22 CME events, the average and 1-$\sigma$ standard deviation of $\alpha_{Bmax}$ is $-1.67 \pm 0.88$, of $\alpha_{Bavg}$ is $-1.60 \pm 0.96$, of $\alpha_r$ based on the radial size without correcting for the radial expansion is $0.83 \pm 1.13$, and of $\alpha_r$ with the expansion correction is $1.05 \pm 1.15$. The larger uncertainties, especially for $\alpha_r$, may be due to the fact that our 22 samples are not enough and/or the evolution of the radial size for each event can be significantly different. For $\alpha_B$, the fitted indices are consistent with past results, e.g., $-1.64$ in \citet{leitner2007}, but slightly lower than that from $-1.8$ to $-1.9$ in \citet{lugaz2020}. For $\alpha_r$ without correcting for the expansion speed, the average value is also close to previous statistical results, e.g., 0.78 to 0.92 \citep{bothmer1998,liu2005}. The average of $\alpha_r$ with the expansion speed correction becomes higher and is close to the result found by \citet{leitner2007} for the CME samples $\le$1~au. Furthermore, as for our samples $\alpha_r \approx -\alpha_B /2$, which indicates the conservation of (axial) magnetic flux inside CMEs during the CME propagation \citep{dumbovic2018}. Table \ref{tab: alpha_index} lists the averages and 1-$\sigma$ uncertainties of $\alpha_r$ and $\alpha_B$ estimated in this paper and obtained from some past studies.

The right panels of Figure~\ref{fig: cmp_exp_para} show the comparisons between $\zeta_{\rm{GCS}}$ and $\alpha_B$ and $\alpha_r$. The maximum and average magnetic fields and the radial size estimates without and with correcting the radial expansion are incorporated. Based on the estimated $\rho_S$ values in Table~\ref{tab: cc}, there is no correlation of neither $\zeta_{\rm{GCS}}$ versus $\alpha_B$ nor $\zeta_{\rm{GCS}}$ versus $\alpha_r$, and $V_{\rm exp}$, $V_{\rm center}$, and $V_{\rm front}$ in the GCS model do not correlate with $\alpha_B$ and $\alpha_r$.

\section{Discussion}\label{sec: dis}
\subsection{Uncertainties in Remote and In-Situ Estimates}
\begin{figure}[!hbt]
	\centering
	\includegraphics[width=0.92\textwidth]{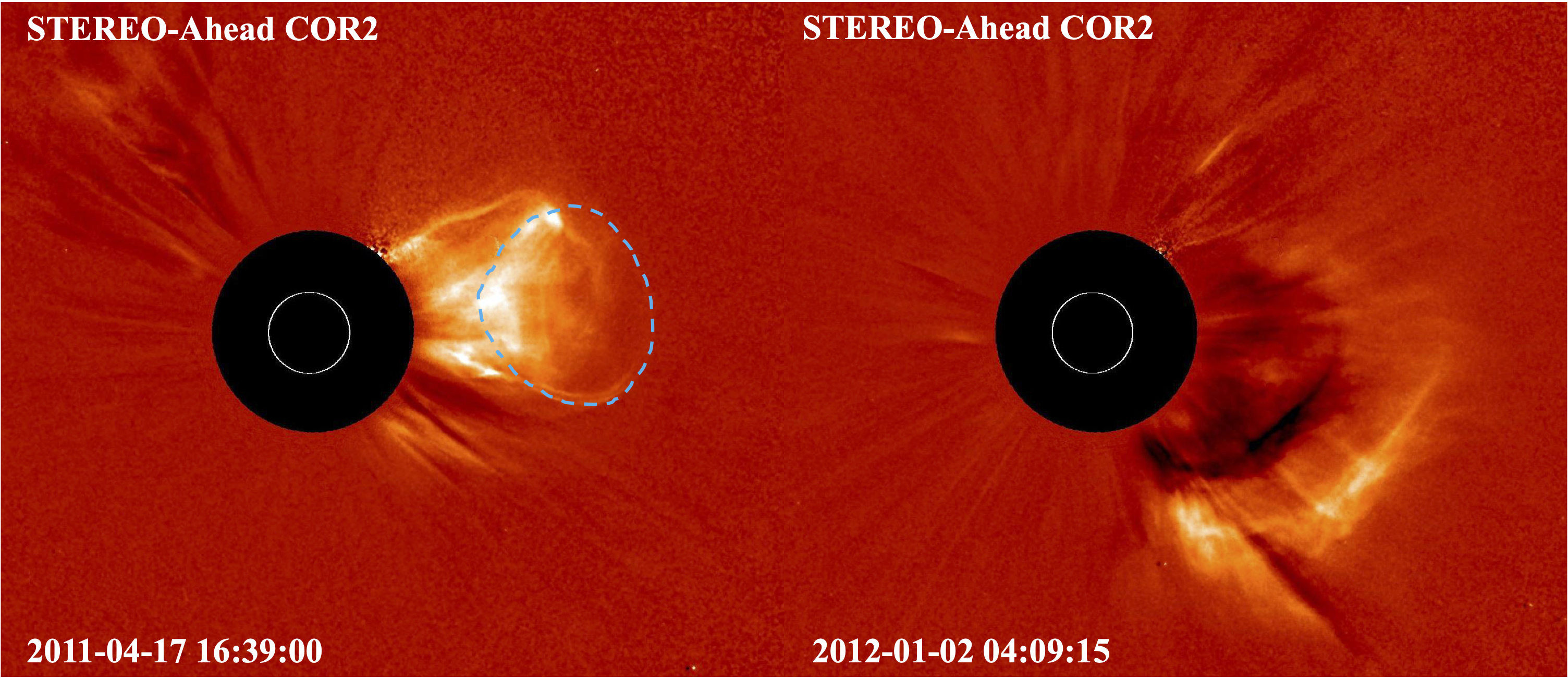}
	\caption{\small Two CME cases illustrating the difficulties in identifying the CME trailing edge in the running-difference coronagraph images. The dashed circle in the left panel outlines the CME boundary.}
	\label{fig: cme_trail_edge}
\end{figure}

The correct and consistent identification of the CME boundaries, not only for remote observations but also for in-situ measurements, influences the derivation of the radial size and expansion parameters significantly. When using the GCS model to study the CME expansion in the corona, identifying the trailing edge of CMEs in the corona without magnetic field measurements may further affect fitting $\kappa$. As compared to the ease of identifying the relatively bright and sharp boundary of the leading edge, the diffusive trailing structure in coronagraph images is difficult to identify. Figure~\ref{fig: cme_trail_edge} shows two opposite cases to illustrate the difficulties in identifying the CME trailing edge in coronagraph images.
As for the CME in the left panel, the V-shape structure at the trailing part and the dark cavity \citep[see more details about CME observational structure in, e.g.,][]{lugaz2012,howard2017,zhuang2022} can help identify the boundary precisely. The blue dashed circle outlines the identified CME boundary. However, the trailing part of the other CME in the right panel is diffuse and hard to determine, and similar observations are not unusual for our cases.

For in-situ measurements, even with solar wind plasma measurements, exact CME boundaries (especially the rear edge) are not clear sometimes \citep[e.g. see discussion in][]{richardson2010,kilpua2013}. This problem is even worse for the observations at MESSENGER and VEX due to (1) the lack of plasma measurements and (2) the influences of planetary magnetosphere crossings. The launched Parker Solar Probe and Solar Orbiter with plasma measurements and moving within 1~au may help with this issue. However, it may take several years for $\sim$20 CMEs to be observed in radial conjunction by two spacecraft including Parker Solar Probe and Solar Orbiter.

For the magnetic field and radial size calculations, we directly use the spacecraft in-situ measurements which only measure a time series along the CME pass path, and we calculate the size based on the product of the CME velocity and duration, which may not well represent the CME global configuration in 3-D space. For example, we need to consider the effects of the distance of the observational path to the axis/center (the impact parameter) of the CME and/or the inclination of the CME axis relative to the Sun-spacecraft line. Various CME fitting methods were developed for a better understanding of the CME global magnetic configuration \citep[see reviews in, e.g.,][]{forbes2006,al-haddad2013,zhang2021}, and help with more precise estimations of the maximum (average) magnetic field strength and/or radial size. \citet{leitner2007} compared the exponential indices of the maximum magnetic field strength by least-squares fitting using a force-free magnetic configuration and by the direct measurements. They found that both $\alpha_B$ and $\alpha_r$ obtained from the fitting are different from those based on direct measurements. In this paper, such fitting techniques are not used, and we note that sometimes these fitting techniques do not reveal true estimates \citep{riley2004,al-haddad2011,al-haddad2019}. We consider that the use of such techniques would also add additional uncertainties in the results of the in-situ analysis rather than providing more insight. However, we determine, from the GCS model, the CME radial size along the Sun--spacecraft line and in the ecliptic plane. Therefore, comparing this directly with the in-situ estimates in the ecliptic plane still ensures a reasonable comparison.

As mentioned before, we assume that the CME propagation direction and self-similar expansion from the GCS model are maintained during the propagation. However, the deflection, both in latitude and longitude \citep{shen2011,kay2015,mostl2015,zhuang2019}, deformation of CMEs, e.g., the ``pancaking effect'' \citep{manchester2004,riley2004,savani2010,vrsnak2019}, interaction between CMEs and other transients \citep{lugaz2012,lugaz2017a,winslow2016,winslow2021b,scolini2021}, and magnetic reconnection between the CME and ambient magnetic field \citep{dasso2006,ruffenach2012,lavraud2014,wang2018,vrsnak2019} may play a significant role in the CME evolution in the corona and interplanetary space, and lead to the inconsistencies of the radial expansion estimated by combining remote observations and in-situ measurements. As for the CME deformation, the fixed $\kappa$ ignores this effect and requires that the CME expands self-similarly. The self-similar expansion may sometimes break \citep{vrsnak2019}. Since, however,  not all of the 22 CMEs can be well tracked in STEREO-HI images, the consideration of a fixed $\kappa$ is the best assumption for the current analysis.

\subsection{Improvement of CME Size Prediction}
\begin{figure}[!hbt]
	\centering
	\includegraphics[width=\textwidth]{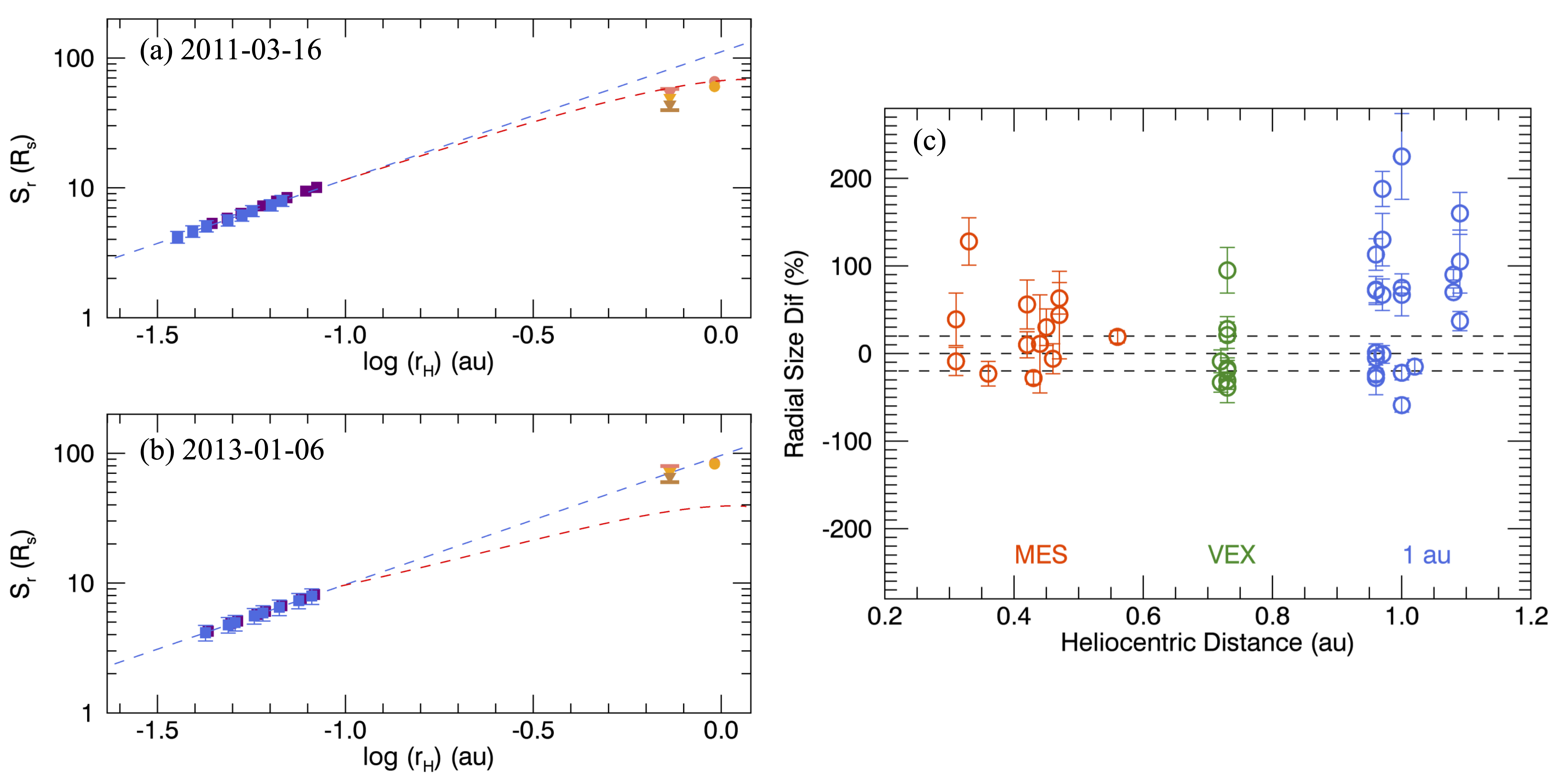}
	\caption{\small Panels (a) and (b): predictions of the CME radial size for two case events obtained by assuming a linearly decreasing CME expansion speed, shown by the dashed red curve and added to the previous results as shown in Figures \ref{fig: radial_exp_p1} and \ref{fig: radial_exp_p3}. Panel (c): difference of the radial sizes associated with their uncertainties} estimated by the prediction with distance-varied expansion speed and by direct in-situ measurements, similar to Figure~\ref{fig: radial_dif} (an outlier of size difference at MESSENGER for the 2011-November-3 event is not shown).
	\label{fig: exp_predict}
\end{figure}

In this section, we discuss attempts to improve the prediction of the CME radial size from coronal remote observations. While the CME size is not as important as the speed or magnetic field strength and orientation for geo-effectiveness, it determines how long the forcing of the magnetosphere by the CME can last and is therefore also important. We first test a method by using a varied expansion speed to extrapolate the CME radial size from the GCS model results. We make the same assumption as in Section~\ref{sec: quanti1} that the expansion speed decreases linearly from the inner boundary of the heliosphere of 20~$R_\odot$ to 1~au. This linearly decreasing expansion speed can result in a better prediction for some cases, as indicated by one example on 2011 March 16 as shown in Figure~\ref{fig: exp_predict}(a), but also leads to poor predictions, e.g., the 2013-January-6 event as shown in Figure~\ref{fig: exp_predict}(b).
In these two panels, dashed red curves, indicating the prediction by using the distance-adjusted expansion speed, are added to the previous comparison results as shown in Figures~\ref{fig: radial_exp_p1} and \ref{fig: radial_exp_p3}. Figure~\ref{fig: exp_predict}(c) shows the difference between the predictions by remote observations and the in-situ estimates at different distances. Even with this time-varied expansion speed, the $\sim 150\%$ outlier at MESSENGER still exists, and this value is neither considered nor plotted. In general, this assumption about the expansion speed results in a larger radial size than the measured one, as indicated by the data points above zero (horizontal line) and the positive averages of the size difference. The mean difference and mean absolute difference of the radial sizes between the GCS and in-situ estimates are listed in Table~\ref{tab: size_dif}. Compared to Figure~\ref{fig: radial_dif}, it shows that the predictions of the radial size at MESSENGER and near 1~au become even worse with larger size differences (also supported by the Wilcoxon test). Therefore, using a linearly decreasing expansion speed does not improve the size prediction.

We recall the $\zeta_{\rm{GCS}}$-$S_r$ correlation in Section \ref{sec: cmp_radial_size}. It indicates that the CME radial size in interplanetary space depends on the initial $\zeta$ when the CME is in the corona. In general, if a CME has both high expansion and propagation speeds, its radial size may not become large because the high propagation speed acts on limiting the CME expansion time. Only a large $\zeta$, roughly indicating a higher proportion of the CME radial expansion speed to the propagation speed, leads to a large CME radial size. As the CME propagates further, its propagation speed gets close to the solar wind speed due to the drag force, and thus the evolution of the CME radial size may primarily depend on the CME expansion speed.

\section{Summary}\label{sec: con}
We investigate the evolution of the radial expansion from the corona to interplanetary space of 22 radially aligned CMEs between the years 2010 and 2013 by combining remote and in-situ observations. The radial alignment occurs between MESSENGER, Venus Express, Wind/ACE, STEREO-A, and STEREO-B at different heliocentric distances with a longitudinal separation of less than $35^\circ$. We use the GCS model based on coronagraph images from multiple viewpoints to estimate the CME expansion behavior in the corona. We compare the radial size extrapolated by the GCS model with that estimated in situ, and find that the evolution of the radial expansion behavior in the corona is not always consistent with that in interplanetary space, especially when the CME is farther away from the Sun. The differences between the radial size obtained from the GCS model extrapolation and in-situ measurements are found to change with distance. Out of these 22 events, only three have a consistent radial size evolution between the remote and two in-situ observations, eight events have a consistency between the remote observations and in-situ measurements at one spacecraft only (two near 1~au, and six at MESSENGER), and 11 for neither of the radially aligned spacecraft.

We focus on different types of radial expansion parameters, including the expansion speed $V_{\rm{exp}}$, a dimensionless parameter $\zeta$ indicating the local expansion for both types of observations, and the indices based on the decrease in the CME magnetic field strength with distance ($\alpha_B$) and the increase in the CME radial size with distance ($\alpha_r$) indicating a global expansion by using in-situ measurements. We find that the correlation between $V_{\rm{exp}}$ in the corona and that near 1~au is moderate, but not for the parameter of $\zeta$. As theoretically defined in the GCS model, $V_{\rm{exp}}$ and $\zeta_{GCS}$ are a function of the fitted parameter $\kappa$, indicating the importance of accurately identifying the CME trailing edge in coronagraphs. The relationships between $V_{\rm{exp}}$ and $\zeta$ in the GCS model versus the radial size $S_r$ estimated in situ are also studied, and moderate correlations between these parameters are found. However, the $V_{\rm{exp}}$-$S_r$ and $\zeta_{\rm GCS}$-$S_r$ correlations might be effective in different heliocentric distance ranges. We further focus on the relationship between $\zeta_{\rm GCS}$ and $\alpha_B$ and $\alpha_r$, and find that there are no such correlations. Overall, our results show that the CME radial expansion behavior may change significantly from the corona to interplanetary space, caused by different mechanisms acting on the expansion processes, which leads to the inconsistency between the CME expansion parameters at different heliocentric distances.

\textbf{Acknowledgement} We acknowledge the use of MESSENGER data, available at Planetary Data System \url{https://pds.nasa.gov}, Venus Express data, available at European Space Agency's Planetary Science Archive \url{https:// www.cosmos.esa.int/web/psa/venus-express}, and NASA/GSFC's Space Physics Data Facility's CDAWeb service for STEREO, Wind, and ACE data, available at \url{https: cdaweb.gsfc.nasa.gov/index.html}. We also thank the use of the data of SOHO/LASCO and STEREO/SECCHI from \url{sdac.virtualsolar.org/cgi/search}. Research for this work was made possible by NASA grant 80NSSC20K0431. N.L. and B.Z. acknowledge the NASA grants 80NSSC17K0009 and 80NSSC19K0831 and the NSF grant AGS-2301382. C.F. acknowledges the NASA grant 80NSSC19K1293. C.S. acknowledges the NASA Living With a Star Jack Eddy Postdoctoral Fellowship Program, administered by UCAR's Cooperative Programs for the Advancement of Earth System Science (CPAESS) under award no.\ NNX16AK22G. 
N.A. acknowledges support from NSF AGS-1954983 and NASA ECIP 80NSSC21K0463.

\begin{table}[!htb]
	\centering
	\caption{CME events observed by the radially aligned spacecraft at different heliocentric distances.}
	\begin{tabular}{|c|c|c|c|c|c|c|c|c|c|}
		\hline
		CME date & Conjunction & \multicolumn{3}{c|}{Remote (GCS)} & \multicolumn{5}{c|}{In-situ (1~au)} \\ \cline{3-10}
		 & & $V_{\rm{exp}}$ & $V_{\rm center}$ & $\zeta$ & $V_{\rm{exp-fit}}$ & $V_{\rm{exp-meas}}$ & $V_{\rm center}$ & $\zeta_{\rm{fit}}$ & $\zeta_{\rm{meas}}$ \\ \hline
		2010-06-02 & VEX-STB & 88 & 395 & 1.22 & 20 & 13 & 371 & 1.10 & 0.68 \\ \hline
		2010-06-16 & MES-Wind & 67 & 358 & 1.20 & 35 & 37 & 365 & 0.57 & 0.60 \\ \hline
		2010-11-03 & MES-STB & 92 & 349 & 1.26 & -20 & -7 & 392 & -0.46 & -0.16 \\ \hline
		2010-12-12 & MES-STA & 56 & 210 & 1.27 & 33 & 18 & 483 & 0.83 & 0.45 \\ \hline
		2011-03-16 & VEX-STA & 202 & 520 & 1.39 & 42 & 32 & 451 & 0.67 & 0.52 \\ \hline
		2011-04-17 & VEX-STA & 45 & 223 & 1.20 & 27 & 21 & 320 & 1.01 & 0.77 \\ \hline
		2011-09-06 & MES-STA & 182 & 597 & 1.30 & -15 & -75 & 388 & -0.16 & -0.76 \\ \hline
		2011-11-03 & MES-VEX-STB & 262 & 937 & 1.28 & 135 & 95 & 514 & 1.50 & 0.95 \\ \hline
		2011-11-17 & VEX-STB & 289 & 816 & 1.35 & 29 & 15 & 506 & 0.43 & 0.22 \\ \hline
		2011-12-29 & MES-STA & 188 & 671 & 1.28 & 4 & 4 & 424 & 0.19 & 0.15 \\ \hline
		2012-01-02 & MES-STA & 260 & 1014 & 1.26 & 171 & 15 & 448 & 4.62 & 0.41 \\ \hline
		2012-06-14 & VEX-Wind & 240 & 665 & 1.36 & 122 & 42 & 467 & 4.91 & 1.70 \\ \hline
		2012-07-12 & MES-ACE & 355 & 1357 & 1.30 & 157 & 119 & 508 & 1.30 & 0.99 \\ \hline
		2012-10-21 & MES-STB & 155 & 552 & 1.28 & -5 & 19 & 377 & -0.11 & 0.38 \\ \hline
		2012-11-10 & VEX-STA & 92 & 381 & 1.24 & 61 & 74 & 417 & 0.82 & 1.00 \\ \hline
		2012-11-20 & MES-ACE & 255 & 646 & 1.40 & 41 & 0 & 388 & 1.21 & 0.01 \\ \hline
		2013-01-06 & VEX-STA & 151 & 539 & 1.28 & 11 & 4 & 453 & 0.29 & 0.10 \\ \hline
		2013-02-14 & VEX-STA & 207 & 852 & 1.24 & -9 & 1 & 402 & -0.27 & 0.02 \\ \hline
		2013-04-20 & MES-STA & 158 & 565 & 1.28 & 2 & 32 & 555 & 0.03 & 0.46 \\ \hline
		2013-07-09 & MES-Wind & 127 & 346 & 1.37 & 55 & 42 & 418 & 0.64 & 0.49 \\ \hline
		2013-08-19 & MES-STA & 327 & 708 & 1.50 & 77 & 33 & 412 & 0.82 & 0.35 \\ \hline
		2013-12-26 & MES-STB & 288 & 884 & 1.37 & 41 & 47 & 412 & 0.53 & 0.61 \\ \hline
	\end{tabular}
        \footnotesize
	\begin{tablenotes}
		\item[1] [1] The table lists the CME eruption date, conjunction spacecraft, $\zeta$, $V_{\rm center}$ and $V_{\rm{exp}}$ in the GCS model, $V_{\rm{exp-fit}}$, $V_{\rm{exp-meas}}$, $V_{\rm center}$, $\zeta_{\rm{fit}}$, and $\zeta_{\rm{meas}}$ estimated by the spacecraft near 1~au. The unit of the speed is km\,s$^{-1}$.
	\end{tablenotes}
	\normalsize
	\label{cme_table}
\end{table}
\clearpage

\begin{table}[!htb]
\caption{Radial sizes and size differences associated with their uncertainties (in brackets) estimated remotely (the GCS model) and in situ.}
\begin{adjustwidth}{.5in}{.5in}
\begin{center}
\begin{tabular}{|c|c|c|c|c|}
\hline
    \multirow{2}{*}{CME Date} & $S_r$ at inner s/c ($R_s$) & \multirow{2}{*}{Size Dif } & $S_r$ at outer s/c ($R_s$) & \multirow{2}{*}{Size Dif} \\
    & (GCS ; in-situ) & & (GCS ; in-situ) & \\ \hline

    2010-06-02 & 46 ($\pm 4$) ; 19 ($\pm 3$) & -59\% ($\pm 7\%$) & 65 ($\pm 6$); 28 ($\pm 1$) & -57\% ($\pm 4\%$) \\ \hline
    2010-06-16 & 38 ($\pm 4$) ; 27 ($\pm 2$) & -29\% ($\pm 9\%$) & 69 ($\pm 7$); 60 ($\pm 2$) & -13\% ($\pm 9\%$) \\ \hline
    2010-11-03 & 42 ($\pm 4$) ; 41 ($\pm 7$) & -2\% ($\pm 18\%$) & 99 ($\pm 9$); 64 ($\pm 2$) & -35\% ($\pm 6\%$) \\ \hline
    2010-12-12 & 15 ($\pm 2$) ; 14 ($\pm 3$) & -5\% ($\pm 21\%$) & 37 ($\pm 7$); 44 ($\pm 2$) & 18\% ($\pm 22\%$) \\ \hline
    2011-03-16 & 82 ($\pm 6$) ; 49 ($\pm 6$) & -41\% ($\pm 9\%$) & 107 ($\pm 8$); 63 ($\pm 3$) & -41\% ($\pm 5\%$) \\ \hline
    2011-04-17 & 53 ($\pm 5$) ; 39 ($\pm 2$) & -26\% ($\pm 8\%$) & 70 ($\pm 7$); 37 ($\pm 2$) & -46\% ($\pm 6\%$) \\ \hline
    2011-09-06 & 31 ($\pm 3$) ; 24 ($\pm 4$) & -23\% ($\pm 14\%$) & 98 ($\pm 8$); 144 ($\pm 1$) & 47\% ($\pm 12\%$) \\ \hline
    \multirow{2}{*}{2011-11-03} & 33 ($\pm 3$) ; 82 ($\pm 13$) & 150\% ($\pm 46\%$) & \multirow{2}{*}{81 ($\pm 7$); 145 ($\pm 4$)} & \multirow{2}{*}{80\% ($\pm 14\%$)} \\
    & 54 ($\pm 5$) ; 31 ($\pm 4$) & -42\% ($\pm 9\%$) & & \\ \hline
    2011-11-17 & 85 ($\pm 6$) ; 64 ($\pm 6$) & -25\% ($\pm 9\%$) & 127 ($\pm 10$); 80 ($\pm 3$) & -37\% ($\pm 5\%$) \\ \hline
    2011-12-29 & 40 ($\pm 4$) ; 42 ($\pm 5$) & 5\% ($\pm 5\%$) & 48 ($\pm 4$); 25 ($\pm 1$) & -48\% ($\pm 4\%$) \\ \hline
    2012-01-02 & 33 ($\pm 3$) ; 34 ($\pm 17$) & 3\% ($\pm 51\%$) & 69 ($\pm 6$); 40 ($\pm 10$) & -42\% ($\pm 15\%$) \\ \hline
    2012-06-14 & 77 ($\pm 6$) ; 37 ($\pm 10$) & -52\% ($\pm 13\%$) & 105 ($\pm 8$); 30 ($\pm 6$) & -71\% ($\pm 6\%$) \\ \hline
    2012-07-12 & 41 ($\pm 3$) ; 42 ($\pm 15$) & 2\% ($\pm 36\%$) & 87 ($\pm 7$); 149 ($\pm 20$) & 72\% ($\pm 25\%$) \\ \hline
    2012-10-21 & 39 ($\pm 3$) ; 18 ($\pm 3$) & -54\% ($\pm 7\%$) & 96 ($\pm 8$); 61 ($\pm 1$) & -36\% ($\pm 6\%$) \\ \hline
    2012-11-10 & 20 ($\pm 3$) ; 29 ($\pm 4$) & 43\% ($\pm 16\%$) & 82 ($\pm 7$); 110 ($\pm 10$) & 34\% ($\pm 16\%$) \\ \hline
    2012-11-20 & 42 ($\pm 2$) ; 41 ($\pm 9$) & -2\% ($\pm 21\%$) & 139 ($\pm 10$); 44 ($\pm 3$) & -68\% ($\pm 3\%$) \\ \hline
    2013-01-06 & 70 ($\pm 6$) ; 70 ($\pm 7$) & 0\% ($\pm 13\%$) & 92 ($\pm 8$); 85 ($\pm 3$) & -8\% ($\pm 8\%$) \\ \hline
    2013-02-14 & 60 ($\pm 5$) ; 24 ($\pm 3$) & -60\% ($\pm 5\%$) & 79 ($\pm 7$); 54 ($\pm 3$) & -32\% ($\pm 2\%$) \\ \hline
    2013-04-20 & 68 ($\pm 6$) ; 53 ($\pm 8$) & -22\% ($\pm 14\%$) & 89 ($\pm 8$); 63 ($\pm 1$) & -29\% ($\pm 6\%$) \\ \hline
    2013-07-09 & 56 ($\pm 3$) ; 48 ($\pm 7$) & -14\% ($\pm 13\%$) & 125 ($\pm 9$); 105 ($\pm 7$) & -16\% ($\pm 8\%$) \\ \hline
    2013-08-19 & 69 ($\pm 5$) ; 82 ($\pm 19$) & 19\% ($\pm 28\%$) & 253 ($\pm 18$); 112 ($\pm 10$) & -55\% ($\pm 6\%$) \\ \hline
    2013-12-26 & 47 ($\pm 3$) ; 29 ($\pm 5$) & -38\% ($\pm 10\%$) & 110 ($\pm 8$); 93 ($\pm 3$) & -15\% ($\pm 7\%$) \\ \hline
\end{tabular}
\end{center}
\end{adjustwidth}
    \footnotesize
    \begin{tablenotes}
        \item[1] [1] The event on 2011 November 3 was observed subsequently by three radially aligned spacecraft. The radial sizes and the size differences at MESSENGER (top) and VEX (bottom) are listed together.
    \end{tablenotes}
    \normalsize
    \label{tab: radial_size_tb}
\end{table}

\begin{table*}[!htb]
\caption{Consistency information of the radial size estimated using remote and in-situ observations.}
\begin{tabular}{|c|c|c|c|}
    \hline
    \multirow{2}{*}{Consistency} & Group I (3) & Group II (8) & Group III (11) \\
    & Two s/c agreement & One s/c agreement & No s/c agreement \\ \hline
    \multirow{8}{*}{Event Date} & & 2010-06-16 (Wind) & \\
    & & 2010-11-03 (MES) & 2010-06-02, 2011-03-16 \\
    & & 2011-12-29 (MES) & 2011-04-17, 2011-09-06 \\
    & 2010-12-12 (MES-STA) & 2012-01-02 (MES) & 2011-11-03, 2011-11-17 \\
    & 2013-01-06 (VEX-STA) & 2012-07-12 (MES) & 2012-06-14, 2012-10-21 \\
    & 2013-07-09 (MES-Wind) & 2012-11-20 (MES) & 2012-11-10, 2013-02-14 \\
    & & 2013-08-19 (MES) & 2013-04-20 \\
    & & 2013-12-26 (STB) & \\ \hline
\end{tabular}
\footnotesize
\begin{tablenotes}
    \item[1] The spacecraft (MES: MESSENGER; VEX: Venus Express; STA: STEREO-A; STB: STEREO-B) for the consistent events is listed in the brackets in the second and third columns.
\end{tablenotes}
\normalsize
\label{tab: consistency}
\end{table*}

\begin{table*}[!htb]
        \centering
	\caption{Mean difference and mean absolute difference of the radial sizes associated with their uncertainties (in brackets)} between the GCS model and in-situ estimations.
	\begin{tabular}{|c|c|c|c|c|}
		\hline
		\multirow{3}{*}{S/C} & \multicolumn{2}{c|}{Mean Dif} & \multicolumn{2}{c|}{Mean absolute Dif} \\ \cline{2-5}
		
		& Constant $V_{\rm{exp}}$ & Decreasing $V_{\rm{exp}}$ & Constant $V_{\rm{exp}}$ & Decreasing $V_{\rm{exp}}$ \\
		
		& Figure \ref{fig: radial_dif} & Figure \ref{fig: exp_predict}c & Figure \ref{fig: radial_dif} & Figure \ref{fig: exp_predict}c \\ \hline
		
		MES & $-12\% \ (\pm 20\%)$ & $25\% \ (\pm 42\%)$ & $16\% \ (\pm 16\%)$ & $35\% \ (\pm 33\%)$ \\ \hline
		
		VEX & $-29\% \ (\pm 33\%)$ & $-4\% \ (\pm 43\%)$ & $39\% \ (\pm 19\%)$ & $32\% \ (\pm 25\%)$ \\ \hline
		
		1~au & $-18\% \ (\pm 43\%)$ & $60\% \ (\pm 75\%)$ & $41\% \ (\pm 20\%)$ & $74\% \ (\pm 61\%)$ \\ \hline
	\end{tabular}
	\normalsize
	\label{tab: size_dif}
\end{table*}

\begin{sidewaystable}[!htb]
        \centering
	\caption{Correlation coefficient of the expansion parameters estimated by using remote and in-situ observations.}
	\footnotesize
	\begin{tabular}{|cc|c|c|c|c|c|c|c|c|c|c|c|c|c|}
		\hline
		& In-situ & \multicolumn{6}{c|}{1~au} & MES & VEX & \multicolumn{4}{c|}{Global Expansion} \\ \cline{3-14}
		
		GCS & & $V_{\rm{exp-fit}}$ & $V_{\rm{exp-meas}}$ & $\zeta_{\rm{fit}}$ & $\zeta_{\rm{meas}}$ & $V_{\rm center}$ & $S_r$ & $S_r$ & $S_r$ & $\alpha_{Bmax}$ & $\alpha_{Bavg}$ & $\alpha_r$ & $\alpha_{r-exp}$ \\ \hline
		
		\multirow{4}{*}{$V_{\rm{exp}}$} & \multirow{2}{*}{$\rho_P$} & - & 0.29 & - & $\sim$0 & {\bf{0.44}} & 0.38 & {\bf{0.49}} & 0.31 & - & - & - & - \\
		
		& & & ($\pm 0.21$) & & & ($\pm 0.20$) & ($\pm 0.18$) & ($\pm 0.19$) & ($\pm 0.29$) & & & & \\ 
		
		& \multirow{2}{*}{$\rho_S$} & {\bf{0.46}} & 0.22 & 0.21 & $\sim$0 & {\bf{0.45}} & 0.35 & {\bf{0.46}} & 0.23 & $\sim$0 & $\sim$0 & -0.12 & $\sim$0 \\
		
		& & ($\pm 0.17$) & ($\pm 0.21$) & ($\pm 0.21$) & & ($\pm 0.21$) & ($\pm 0.20$) & ($\pm 0.24$) & ($\pm 0.36$) & & & ($\pm 0.22$) &  \\ \hline
		
		\multirow{4}{*}{$\zeta$} & \multirow{2}{*}{$\rho_P$} & - & $\sim$0 & - & $\sim$0 & 0.21 & 0.23 & {\bf{0.46}} & {\bf 0.52} & - & - & - & - \\
		
		& & & & & & ($\pm 0.19$) & ($\pm 0.17$) & ($\pm 0.28$) & ($\pm 0.22$) & & & & \\ 
		
		& \multirow{2}{*}{$\rho_S$} & 0.32 & 0.16 & $\sim$0 & -0.16 & 0.33 & 0.36 & 0.38 & {\bf 0.52} & -0.18 & $\sim$0 & -0.21 & -0.18 \\
		
		& & ($\pm 0.17$) & ($\pm 0.21$) & & ($\pm 0.20$) & ($\pm 0.22$) & ($\pm 0.19$) & ($\pm 0.23$) & ($\pm 0.29$) & ($\pm 0.25$) & & ($\pm 0.22$) & ($\pm 0.22$) \\ \hline		
		
		\multirow{4}{*}{$V_{\rm center}$} & \multirow{2}{*}{$\rho_P$} & - & 0.38 & - & $\sim$0 & {\bf 0.46} & 0.39 & 0.33 & $\sim$0 & - & - & - & - \\
		
		& & & ($\pm 0.26$) & & & ($\pm 0.18$) & ($\pm 0.22$) & ($\pm 0.20$) & & & & & \\ 
		
		& \multirow{2}{*}{$\rho_S$} & 0.36 & 0.15 & 0.19 & $\sim$0 & 0.39 & 0.22 & 0.36 & $\sim$0 & 0.19 & 0.19 & $\sim$0 & $\sim$0 \\
		
		& & ($\pm 0.22$) & ($\pm 0.24$) & ($\pm 0.23$) & & ($\pm 0.20$) & ($\pm 0.22$) & ($\pm 0.26$) & & ($\pm 0.23$) & ($\pm 0.22$) & & \\ \hline		
		
		\multirow{4}{*}{$V_{\rm front}$} & \multirow{2}{*}{$\rho_P$} & - & 0.36 & - & $\sim$0 & {\bf 0.47} & 0.39 & 0.37 & 0.14 & - & - & - & - \\
		
		& & & ($\pm 0.25$) & & & ($\pm 0.19$) & ($\pm 0.21$) & ($\pm 0.19$) & ($\pm 0.31$) & & & & \\
		
		& \multirow{2}{*}{$\rho_S$} & {\bf 0.42} & 0.18 & 0.24 & $\sim$0 & {\bf 0.42} & 0.25 & {\bf{0.40}} & $\sim$0 & 0.17 & 0.18 & $\sim$0 & $\sim$0 \\
		
		& & ($\pm 0.21$) & ($\pm 0.23$) & ($\pm 0.23$) & & ($\pm 0.20$) & ($\pm 0.22$) & ($\pm 0.22$) & & ($\pm 0.23$) & ($\pm 0.22$) & & \\ \hline		

	\end{tabular}
	\begin{tablenotes}
		\item[1] [1] $\rho_P$, $\rho_S$, and the corresponding 1-$\sigma$ uncertainties (in brackets) are listed.
		\item[2] [2] The GCS model parameters include the radial expansion speed $V_{\rm{exp}}$, center speed $V_{\rm center}$, front speed $V_{\rm front}$, and $\zeta$. The in-situ parameters include $V_{\rm{exp-fit}}$, $V_{\rm{exp-meas}}$, $\zeta_{\rm{fit}}$, $\zeta_{\rm{meas}}$, and $V_{\rm center}$ estimated near 1~au, and the radial size by averaging the values with and without correcting for the expansion at MESSENGER, VEX, and the spacecraft near 1~au. Parameters of $\alpha_{Bmax}$, $\alpha_{Bavg}$, $\alpha_r$ (without the expansion corrected), and $\alpha_{r-exp}$ (with the expansion corrected) are also listed.
            \item[3] [3] If the parameter distribution does not follow a normal distribution, then the Pearson correlation coefficient is marked as a dash.
		\item[4] [4] The $\ge$0.4 correlation coefficient is in bold font.
	\end{tablenotes}
	\normalsize
	\label{tab: cc}
\end{sidewaystable}

\begin{table*}[!htb]
        \centering
	\caption{Averages and 1-$\sigma$ standard deviations of $\alpha_r$, $\alpha_{Bmax}$, and $\alpha_{Bavg}$ estimated in this paper and obtained from some past studies.}
	\footnotesize
	\begin{tabular}{|c|c|c|}
		\hline
		Quantity & Average $\pm \sigma$ & Past results \\ \hline
		 & & $0.78 \pm 0.10$ \citep{bothmer1998} \\
		 & & $0.92 \pm 0.07$ \citep{liu2005} \\
		 $\alpha_r$ & $0.83 \pm 1.13$ (without correcting for $V_{\rm{exp}}$) & $0.61 \pm 0.09$ \citep{leitner2007} \\
		 & $1.05 \pm 1.15$ (with correcting for $V_{\rm{exp}}$) & $1.14 \pm 0.44$ \citep{leitner2007} \\
		 & & $0.78 \pm 0.12$ \citep{gulisano2010} \\ \hline
		 
		 & & $-1.67 \pm 0.40$ \citep{leitner2007} \\
		 $\alpha_{Bmax}$ & $-1.64 \pm 0.88$ & $-1.34 \pm 0.71$ \citep{good2019} \\
		 & & $-1.81 \pm 0.84$ \citep{lugaz2020} \\ \hline
		 
		 & & $-1.40 \pm 0.08$ \citep{liu2005} \\
		 $\alpha_{Bavg}$ & $-1.60 \pm 0.96$ & $-1.85 \pm 0.07$ \citep{gulisano2010} \\
		 & & $-1.95 \pm 0.19$ \citep{winslow2015} \\
		 & & $-1.91 \pm 0.85$ \citep{lugaz2020} \\ \hline
	\end{tabular}
	\normalsize
	\label{tab: alpha_index}
\end{table*}
\clearpage

\bibliography{sample631}{}
\bibliographystyle{aasjournal}



\end{document}